\documentclass[journal,twoside,web]{ieeecolor}
\usepackage{cite}
\usepackage{amsmath,amssymb,amsfonts,mathtools}
\usepackage{graphicx}
\usepackage{subcaption}
\usepackage{epstopdf}

\newtheorem{theorem}{Theorem}
\newtheorem{assumption}{Assumption}
\newtheorem{proposition}{Proposition}
\newtheorem{lemma}{Lemma}
\newtheorem{remark}{Remark}
\newtheorem{cor}{Corollary}
\newtheorem{definition}{Definition}
\newtheorem{example}{Example}

\newcommand{\vect}[1]{\boldsymbol{#1}}
\newcommand{\mat}[1]{\boldsymbol{#1}}
\renewcommand{\eqref}[1]{Eq.~(\ref{#1})}  

\def\logoname{LOGO-generic-web}
\definecolor{subsectioncolor}{rgb}{0,0.541,0.855}
\def\journalname{Generic Colorized Journal}

\usepackage{pgf,pgfplots,tikz,amsmath,amsfonts,amssymb}
\pgfplotsset{ 
  compat=newest, 
  label style = {font=\small},
every tick label/.append style={font=\small}
  }
  
\def\BibTeX{{\rm B\kern-.05em{\sc i\kern-.025em b}\kern-.08em
    T\kern-.1667em\lower.7ex\hbox{E}\kern-.125emX}}
\begin{document}

\title{Coevolutionary Dynamics of Actions and Opinions in Social Networks}

\author{Hassan Dehghani Aghbolagh, Mengbin Ye, Lorenzo Zino,  Zhiyong Chen, and Ming Cao
\thanks{H. D. Aghbolagh and Z. Chen are with the School of Engineering, University of Newcastle, Callaghan, 2308, NSW, Australia. Emails: \texttt{hassan.dehghaniaghbolagh@uon.edu.au,} and \texttt{zhiyong.chen@newcastle.edu.au}. M. Ye is with the Centre for Optimisation and Decision Science, Curtin University, Bentley 6102, WA, Australia. Email: \texttt{mengbin.ye@curtin.edu.au}. M. Cao is with the Faculty of Science and Engineering, ENTEG, University of Groningen, Groningen 9747 AG, Netherlands. Email: \texttt{m.cao@rug.nl}. L. Zino was with the Faculty of Science and Engineering, ENTEG, University of Groningen, Groningen 9747 AG, Netherlands; he is now with the Department of Elettronics and Telecommunications, Politecnico di Torino, 10129 Turin, Italy. Email: \texttt{lorenzo.zino@polito.it}.  
The work of H.D. Aghbolagh is supported by an Australian Government Research Training Program (RTP) Scholarship; the work of M. Ye by the Western Australian Government through the Premier's Science Fellowship Program and the Defence Science Centre; the work of L. Zino and M. Cao by the European Research Council (ERC-CoG-771687). }}
\maketitle

\begin{abstract}
Empirical studies suggest a deep intertwining between opinion formation and decision-making processes, but these have been treated as separate problems in the study of dynamical models for social networks. In this paper, we bridge the gap in the literature by proposing a novel coevolutionary model, in which each individual selects an action from a binary set and has an opinion on which action they prefer. Actions and opinions coevolve on a two-layer network. For homogeneous parameters, undirected networks, and under reasonable assumptions on the  asynchronous updating mechanics, we prove that the coevolutionary dynamics is an ordinal potential game, enabling analysis via potential game theory. Specifically, we establish global convergence to the Nash equilibria of the game, proving that actions converge in a finite number of time steps, while opinions converge asymptotically. Next, we provide sufficient conditions for the existence of, and convergence to, polarized equilibria, whereby the population splits into two communities, each selecting and supporting one of the actions. Finally, we use simulations to examine the social psychological phenomenon of pluralistic ignorance.
\end{abstract}

\begin{IEEEkeywords}
dynamics on networks, opinion dynamics, decision making, evolutionary game theory, polarization
\end{IEEEkeywords}

\section{Introduction}
\label{sec:introduction}
\IEEEPARstart{T}{hroughout} the past decades, researchers in different fields have used mathematical models to study different complex dynamical processes in social networks, such as opinion formation and collective decision making~\cite{proskurnikov2017tutorial,riehl2018survey,friedkin2011social_book,granovetter1978threshold,flache2017opdyn_survey}. Development and analysis of such models have helped to understand and predict how individual interactions can lead to complex collective behavior emerging in the wider social network~\cite{friedkin2015problem,Velarde2020Polarization,granovetter1978threshold,peytonyoung2015social_norms,young2011dynamics}.

The field of opinion dynamics studies models of the formation and evolution of opinions in groups of individuals, with roots in the 1950s (see~\cite{proskurnikov2017tutorial,proskurnikov2018tutorial,anderson2019IJAC} for a review). The seminal French-DeGroot model~\cite{french1956_socialpower} established that, under mild network connectivity assumptions, the opinions of all individuals reach a common value (a consensus) due to social influence as captured by a weighted averaging procedure~\cite{degroot1974reaching,french1956_socialpower}. Several generalizations of this model have been proposed. We mention the work by Friedkin and Johnsen~\cite{friedkin1990social}, which posits that an individual's existing prejudices can lead to persistent disagreement, which has been extended and validated experimentally~\cite{friedkin2011social_book,Parsegov2017Multi}. 

On another research front, the study of collective decision-making in networks of interacting agents has gained popularity. In particular, mathematical models based on evolutionary game theory have been used to describe and predict how an individual decides on which action to take (e.g., a product to buy, or a social convention to adopt), choosing from a finite set of possibilities, and how these decisions are revised after interacting with other individuals~\cite{Jackson2015,young2011dynamics,montanari2010spread_innovation,riehl2018survey}. The study of these models, in particular those based on the paradigm of network coordination games and its extensions~\cite{Jackson2015}, has allowed to shed light on the emergent behavior of complex social system, including the evolution of social norms and conventions, and nontrivial interactions between heterogeneous types of agents\cite{ramazi2016networks,peytonyoung2015social_norms}, and has paved the way for the design of intervention policies to control the network~\cite{Riehl2016,como2022targeting}.

Results from the social psychology literature, including empirical evidence, indicate a strong relationship between  the opinions held and decisions taken by individuals. These studies suggest that, within a community, the exchange of opinions can affect individuals' actions, while the way people form opinions can be influenced by observing the actions of others.  For instance, teachers may decide to revise content delivery methods after forming opinions about which method is more effective~\cite{henry2003beyond,patton2003utilization}. In another example,~\cite{greenberg2000dissemination} showed that policymakers shifted their attitudes  about work-oriented reforms when they observed the impact of these actions in
other regions. See~\cite{haidt2001emotional,gavrilets2017collective,lindstrom2018role} for more examples along these lines. Despite such a strong relationship, we also point out that actions and opinions are not always aligned in the real-world. A classic example is the phenomenon of unpopular norms~\cite{centola2005emperor,willer2009false_enforcement,Smerdon2019}, in which a community keeps exhibiting a collective behavior that is disapproved by the most of its members. This is often caused by pluralistic ignorance (see, for instance, the study on alcohol abuse by undergraduates in Princeton University in the 1990s~\cite{prentice1993pluralistic}). In summary, the mechanisms governing the evolution of actions and opinions in communities appear to be nontrivially intertwined. 

Despite evidence of such a complex intertwining, few existing mathematical frameworks are able to capture the coevolution of actions and opinions. Building on the seminal work by Martins~\cite{Martins2008coda},
continuous-opinion discrete-action (CODA) models have been developed and studied~\cite{Chowdhury2016coda,Ceragioli2018quantized,Tang2021coda}, and utilized to study real-world problems such as innovation diffusion~\cite{Martins2009innovation} and competition in duopolies~\cite{Varma2017coda}. CODA models are able to reproduce certain interesting real-world phenomena, including polarization and oscillations~\cite{Chowdhury2016coda,Ceragioli2018quantized}. However, they rely on the simplifying assumptions that i) an individual's action is a direct quantization of their opinion, and ii) only actions are observable in the population. In neglecting more complex social psychological mechanisms that characterize human decision-making, CODA models are unable to capture important phenomena such as unpopular norms or pluralistic ignorance~\cite{centola2005emperor,willer2009false_enforcement,Smerdon2019,prentice1993pluralistic}. Recently, models were proposed to consider coevolution of private and expressed opinions~\cite{duggins2017psychologically,ye2019influence}, but no decision-making process is present. On the other hand,~\cite{centola2005emperor} proposes a decision-making process influenced by the opinions of individuals, which are however assumed to be time-invariant. Some preliminary efforts to model the coevolution of individuals' actions and opinions can be found in our works of~\cite{zino2020two,zino2020coevolutionary}. However, the complexity of the framework proposed in~\cite{zino2020two} hinders its analytical tractability, with findings mostly limited to numerical simulations.

In this paper, we fill in this gap by proposing a novel coevolutionary model in which individuals make decisions and revise their opinions by interacting on a two-layer network, whereby they observe others' actions on an influence layer and share their opinions on a communication layer. The proposed framework is formalized within a game-theoretic paradigm, and unifies the fields of opinion dynamics and collective decision making, by encompassing and generalizing the French-DeGroot~\cite{french1956_socialpower,degroot1974reaching} and the Friedkin-Johnsen~\cite{friedkin1990social,friedkin2011social_book} opinion dynamics models, and network coordination games~\cite{Jackson2015,ramazi2016networks,young2011dynamics,montanari2010spread_innovation}. Specifically, individuals simultaneously update their opinion and action, aiming to maximize a payoff function that consists of four terms, accounting for i) the individuals' tendency to coordinate with others' actions observed on the influence layer; ii) the opinion formation and exchange on the communication layer, iii) the (possible) presence of individuals' existing prejudices, and iv) the individuals' tendency to act consistently with their opinion, respectively. The first term comes from network coordination games~\cite{Jackson2015}, while the second and third come from the Friedkin-Johnsen model~\cite{friedkin1990social}. The fourth term, instead, couples an individual's action with their opinion, ensuring that an individual's decision-making and opinion formation processes are inherently intertwined in a nontrivial fashion as motivated in the above. 

We perform a rigorous analysis of the proposed coevolutionary model leveraging potential game theory. Under  the assumptions that each layer of the network is undirected  and the agent parameters are homogeneous, we prove that the game defining the model is a generalized ordinal potential game~\cite{monderer1996potential}. Exploiting the corresponding potential function, we provide a comprehensive characterization of the long-term behavior of the coevolutionary dynamics, proving global convergence to the Nash equilibria of the coevolutionary game. In particular, we prove that individuals' actions converge in finite time, while opinions converge asymptotically. Our approach follows the works in~\cite{Marden2009game_consensus,Ghaderi2014opinion}, which interpreted opinion dynamics as best-response strategies to a game, but novel analysis was required as our state space is the Cartesian product of discrete (action) and continuous (opinion) sets. Our general convergence result significantly advances the limited existing theoretical analysis of coevolutionary models, that only considered the special scenario in which the opinion formation process is not affected by the decision-making one~\cite{zino2020coevolutionary,zino2020two}.

Then, we use our model to explore two real-world phenomena: \emph{polarization} and \emph{pluralistic ignorance}. Building on our general convergence result, we derive sufficient conditions for the existence of and convergence to polarized equilibria, in which the network is divided into two groups supporting and taking opposite actions. Interestingly, polarization ---often observed in real-world social networks~\cite{Fiorina2008,Cinelli2021}--- occurs in our model without antagonistic interactions or strongly biased assimilation~\cite{altafini2012consensus,xia2015structural_opinions, dandekar2013biased_degroot,xia2020bias} (which are not always realistic assumptions~\cite{Takacs2016}), and without simplified quantized decision-making of CODA models~\cite{Martins2008coda,Chowdhury2016coda,Ceragioli2018quantized}. Finally, we explore the social psychological phenomenon of pluralistic ignorance, which arises when some individuals in a community hold an incorrect assumption about the thoughts, feelings, and/or behaviors of others in the population~\cite{merton1968social_book}. Via simulations, we show that our model is capable of qualitatively reproducing the classical empirical findings in~\cite{prentice1993pluralistic}.

The rest of the paper is organized as follows. We conclude this section with preliminaries including graph theory and game theory. Section~\ref{sec:classic} presents classic models of decision-making and opinion dynamics in a game-theoretic framework. Then, the coevolutionary model is presented in Section~\ref{sec:model}, for which convergence results are provided in Section~\ref{sec:analysis}.  Section~\ref{sec:phenomena} is devoted to the analysis of polarization and pluralistic ignorance phenomena. Finally, Section~\ref{sec:conclusions} concludes the paper.

\subsection{Notation}

We denote the set of real,  nonnegative real, and  nonnegative integer numbers as $\mathbb R$, $\mathbb R_+$, and $\mathbb Z_+$, respectively.  A vector $\vect x$ is denoted with bold lowercase fonts, with $i$th entry $x_i$. A matrix $\vect A$ is denoted with bold capital fonts, where $a_{ij}$ denotes the generic $j$th entry of the $i$th row. The column vector of all ones is denoted by $\vect{1}$ and the identity matrix is $\mat I$, with the appropriate dimension determined in the context.  Given a vector $\vect{x}$ or a matrix $\mat{A}$, $\vect{x}^\top$ and $\mat{A}^\top$ are the transpose vector and matrix, respectively. Given a scalar quantity $x\in\mathbb R$, the function $\text{sgn}(x)$ denotes its sign, with $\text{sgn}(0)=0$.

\subsection{Graph Theory}

A (weighted) directed network is defined as a triple $\mathcal G_A=(\mathcal V,\mathcal E_A,\mat{A})$, where $\mathcal V=\{1,\dots,n\}$ is the \emph{node set}, indexed by positive integer numbers; $\mathcal E_A\subseteq \mathcal V\times\mathcal V$ is the \emph{edge set}, where $(i,j)\in\mathcal E_A$ if and only if there is a link from node $i$ to node $j$; and $\mat{A}\in\mathbb R_{+}^{n\times n}$ is the \emph{weight matrix}, so that its generic entry $a_{ij}>0\iff(i,j)\in\mathcal V$. The edge $(i,i)\in\mathcal E_A$ represents a \emph{self-loop}. A network is undirected iff $\mat A = \mat A^\top$. We say that a weight matrix is \emph{row-stochastic} (\emph{column-stochastic}) if its rows (columns) sum to $1$, that is, if $\mat{A}\vect{1}=\vect{1}$ ($\vect{1}^\top\mat{A}=\vect{1}^\top$). Weight matrices that are both row- and column-stochastic are said to be \emph{doubly-stochastic}. 

Given a network $\mathcal G_A=(\mathcal V,\mathcal E_A,\mat{A})$ and a node $i\in\mathcal V$, we define the neighbors of node $i$ as the set $\mathcal N^A_i:=\{j:(i,j)\in\mathcal E_A\}$. Given a pair of nodes $i$ and $j$, we say that $i$ is connected with $j$ if there exists a sequence of edges $(e_1,\dots,e_k)$ such that $e_h\in\mathcal E_A$, for all $h=1,\dots,k$, $e_1=(i,p)$, and $e_k=(q,j)$, for some $p,q\in\mathcal{V}$. We say that a network is (strongly) \emph{connected} if and only if $i$ is connected with $j$, for every pair of distinct nodes $i,j\in\mathcal V$. Moreover, $\mathcal{G}_A$ is strongly connected if and only if $\mat A$ is an irreducible matrix~\cite{godsil2001algebraic}.

\begin{figure}
    \centering
   \includegraphics[scale=0.9]{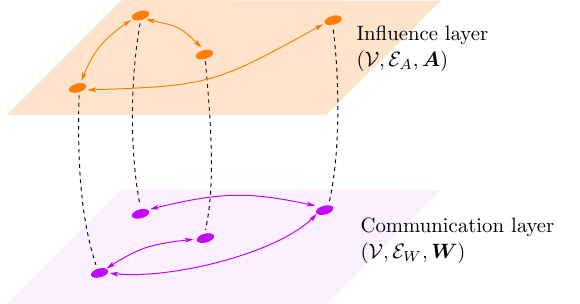}
    \caption{The two-layer network structure utilized in the model.\vspace{-10pt} }
    \label{fig:network}
\end{figure}

Given two networks $\mathcal G_A$ and $\mathcal G_W$ with the same node set $\mathcal{V}$, we define a \emph{two-layer network} as $\mathcal G=(\mathcal V,\mathcal E_A,\mat{A},\mathcal E_W,\mat{W})$, where the edge set and weight matrix of each original network form a layer of the two-layer network~\cite{Boccaletti2014}. In general, $\mathcal{E}_A \neq \mathcal{E}_W$ and thus the two layers have different edges. We say that the two-layer network is connected if and only if both layers are separately connected. An example of a two-layer network is illustrated in Fig.~\ref{fig:network}.

\subsection{Game Theory}

A game $\gamma=(\mathcal{V},\mathcal{A},\vect{f})$ is defined by a set of players $\mathcal{V}=\{1,\ldots,n\}$; a set of strategies $\mathcal{A}$ that the players can choose, where a strategy $\zeta\in\mathcal{A}$ can be, in general, a scalar or a vector\footnote{A more general formulation, which is beyond the scope of this paper, may consider scenarios in which the strategy set can vary between players.}; and a payoff vector, $\vect{f}=[f_1,\dots,f_n]^\top$, where $f_i:\mathcal{A}\to \mathbb{R}$ is the \emph{payoff function} of player $i\in\mathcal V$. Let ${z_i}\in\mathcal{A}$ represent the strategy of individual $i$; then, the state variable of the game $\vect z\in\mathcal{A}^n$ can be defined by gathering the strategies of the $n$ individuals into a single vector (or matrix if strategies are vectors). The payoff function $f_i(\zeta,\vect{z})$ determines the payoff that individual $i$ receives for choosing strategy $\zeta\in\mathcal A$, given the state of the system is  $\vect{z}$. In the following, we will provide the definition of a generalized ordinal potential game~\cite{monderer1996potential}, which will be used in the rest of this paper.\smallskip

\begin{definition}
A game $\gamma=(\mathcal{V},\mathcal{A},\vect{f})$ is called a \emph{generalized ordinal potential} game if there exist a potential function $\Phi:\mathcal{A}\to\mathbb R$ such that for any pair of states $\vect{z}$ and $\vect{z'}$ that differs in a single component ${z_i}\neq {z_i}'$ (i.e., the state of a single player), then, for all $i\in \mathcal V$, the following holds:
\begin{equation} \label{eq:gopg_ineq}
f_i({z_i},\vect{z})-f_i(z_i',\vect{z})>0\implies \Phi(\vect{z})-\Phi(\vect{z}')>0.
\end{equation} 
\end{definition}\smallskip

Notice that in some game-theoretic literature, when referring to changes in the potential due to updates in the state of a player $i$, the potential function is written as a function of two arguments, that is, $z_i$ and a vector containing all the other entries of the state $\vect z$. For the sake of readability, we decided to adopt a simpler notation. 

Finally, suppose that at each time step $t\in \mathbb Z_+$, individual $i$ has the opportunity to revise their state. In this paper, we will consider individuals using a well-known concept in game theory known as best-response updating to revise their state. Formally, we can define the set of best responses as follows.\smallskip

\begin{definition}
Given a game $\gamma=(\mathcal{V},\mathcal{A},\vect{f})$, the \emph{best-response strategies} for player $i\in\mathcal{V}$ given the state of the system $\vect{x}$ are defined as
\begin{equation} \label{eq:best_response_generic}
\mathcal{B}_i\big(f_i(\cdot,\vect{z})\big):=\rm{argmax}_{ \vect{\zeta}\in\mathcal{A}} f_i({\zeta},\vect{z}).
\end{equation} 
\end{definition}

Note that, in general, the best response is not unique, and multiple strategies might be contained in the set $\mathcal{B}_i\big(f_i(\cdot,\vect{z})\big)$.

\section{Decision-Making and Opinion Dynamics}\label{sec:classic}

Before proposing our coevolutionary model, we briefly illustrate how decision-making and opinion dynamics have been cast, independently, as best-response dynamics in a game-theoretic formalism, and we discuss the motivations for developing a coevolutionary framework.

\subsection{Decision-Making as a Network Coordination Game}\label{sec:coord_game}

In the last decade, game theory has become increasingly popular to model collective decision-making in communities, especially through the lens of network coordination games~\cite{montanari2010spread_innovation,young2011dynamics,Jackson2015,ramazi2016networks}. These models consider a population of $n \geq 2$ individuals, each assigned a binary variable $x_i(t)\in \{-1, 1\}$, representing individual $i$'s \emph{action} at discrete-time instants $t \in \mathbb Z_+$. Actions are gathered in the \emph{action vector} $\vect{x}(t) = [x_1(t), \hdots, x_n(t)]^\top \in \{-1,+1\}^n$, which is the network's state. Individuals observe others' actions on a weighted network $\mathcal{G}_A=(\mathcal V,\mathcal E_A, \mat A)$, with the row stochastic weight matrix $\mat A$. 

Each individual engages in a pairwise coordination game with each and every neighbor, with unit payoff for coordinating on action $-1$ and payoff of $1+\alpha$ for coordinating on action $1$, where $\alpha \geq 0$. The scenario $\alpha > 0$ describes action $+1$ having an advantage over $-1$, e.g. for innovations~\cite{montanari2010spread_innovation,young2011dynamics}. 
Given state $\vect{x}$, each individual $i$ is thus associated with the following payoff function for selecting action $\zeta_a\in\{-1,+1\}$:
\begin{equation}\label{eq:f_function_action}
f^a_i(\zeta_a,\vect{x})=\frac14\sum_{ j\in \mathcal{V}}a_{ij}[(1-x_j)(1-\zeta_a)+(1+\alpha)(1+x_j)(1+\zeta_a)].
\end{equation}

Observe that the best-response strategy for such a payoff function yields a threshold such that $+1\in \mathcal{B}_i(f^a_i(\cdot,\vect{x}))$ if and only if a discriminant quantity $\delta^a(\vect x)$ satisfies
\begin{equation}
   \delta^a(\vect{x})= \sum\nolimits_{j\in\mathcal V}a_{ij}\big[2x_j+\alpha(1+x_j)\big]\geq 0.
\end{equation}
In other words, action $+1$ is a best response if and only if a sufficiently large fraction of the neighbors of $i$ adopt it (weighted by the matrix $\mat{A}$ and the advantage $\alpha$). Characterizing the equilibria and studying the long-term behavior of a population of individuals who update their action following a best-response strategy in a network coordination game has been extensively studied~\cite{morris2000contagion,Jackson2015}, even for heterogeneous agents~\cite{ramazi2016networks} and in the presence of noise~\cite{montanari2010spread_innovation,young2011dynamics}.

\subsection{Friedkin-Johnsen Opinion Dynamics Model}~\label{sec:FJ}

In the Friedkin-Johnsen model~\cite{friedkin1990social,friedkin2011social_book}, each individual $i$ of a population of $n \geq 2$ individuals is assigned a continuous variable $y_i(t)\in [-1, 1]$, representing the individual's \emph{opinion} at discrete-time instant $t \in \mathbb Z_+$. Opinions are gathered in the \emph{opinion vector} $\vect{y}(t) = [y_1(t), \hdots, y_n(t)]^\top \in [-1,+1]^n$, which is the network's state. Individuals share their opinions through interactions on a weighted network $\mathcal{G}_W=(\mathcal V,\mathcal E_W, \mat W)$, with the row stochastic weight matrix $\mat W$.

Specifically, individual $i$'s action at the next time step $y_i(t+1)$ is obtained by averaging the opinions shared by the neighboring individuals and individual $i$'s existing prejudice $u_i\in[-1,1]$. The parameter $\gamma_i\in[0,1]$ provides a weighting that captures individual $i$'s level of attachment to their prejudice relative to the influence from opinions of their neighbors. Thus, the update rule reads
\begin{equation}\label{eq:opinion}
    y_i(t+1)=(1-\gamma_i)\sum\nolimits_{j\in\mathcal{V}}w_{ij}y_j(t)+\gamma_iu_i.
\end{equation}

Following~\cite{Marden2009game_consensus,Ghaderi2014opinion}, the Friedkin-Johnsen model can be equivalently cast in a game-theoretic framework. In fact, with payoff the function
\begin{equation}\label{eq:f_function_opinion}
f^o_i(\zeta_o,\vect{y})=-\frac12(1-\gamma_i)\sum_{j\in\mathcal{V}}w_{ij}(\zeta_o-y_j)^2-\frac12\gamma_i(\zeta_o-u_i)^2
\end{equation}
the update rule in \eqref{eq:opinion} can be interpreted as a best-response strategy, that is, $y_i(t+1)=\mathcal{B}_i(f^o_i(\cdot,\vect{y}))$.

\subsection{Motivation for a Coevolutionary Framework}

As discussed in the Introduction, there is strong evidence from the social psychology literature that an individual's actions and those of their neighbors can influence the individual's opinion, and vice versa~\cite{patton2003utilization,greenberg2000dissemination,henry2003beyond,haidt2001emotional,gavrilets2017collective,lindstrom2018role}. This major body of empirical evidence suggests that decoupling the decision making and opinion formation and studying them one independently of the other may be an oversimplification, which restricts the range of behaviors that such models can reproduce and predict. 

Surprisingly, few efforts have explored this coupling via mathematical models except our two related works~\cite{zino2020two,zino2020coevolutionary}, which explored a scenario in which the opinion formation and decision-making processes are indeed intertwined, but the two update rules are distinct rather than obtained from a single payoff function, and analytical results are limited to the scenario in which the former is not affected by the latter. In the following, instead, we will propose a coevolutionary framework, in which individuals simultaneously make decisions and revise their opinions, and we illustrate how important real-world collective population phenomena could arise from the coevolutionary framework, namely polarization and pluralistic ignorance.

\section{Coevolutionary Model}\label{sec:model}

In this section, we propose the coevolutionary model, which couples the network coordination games and opinion dynamics models within a single game-theoretic framework.

\subsection{Setting}

We consider a population of $n \geq 2$ interacting individuals, each assigned a two-dimensional state variable (or strategy) $\vect{z}_i(t)=(x_i(t), y_i(t)) \in\mathcal{A}= \{-1, 1\}\times [-1, 1]$, representing individual $i$'s \emph{action} and \emph{opinion}, respectively, at discrete-time instant $t \in \mathbb Z_+$. Individual~$i$'s action at time $t$,  $x_i(t)$, represents their choice on two alternative actions of $-1$ and $+1$, while the opinion $y_i(t)\in[-1,1]$, represents individual~$i$'s attitude orientation towards the two actions. Actions are gathered in the \emph{action vector} $\vect{x}(t) = [x_1(t), \hdots, x_n(t)]^\top \in \{-1,+1\}^n$, while opinions are gathered in the \emph{opinion vector} $\vect{y}(t) = [y_1 (t), \hdots, y_n(t)]^\top \in [-1,1]^n$. These two vectors determine the \emph{state of the system}, which can be written by gathering them in an $n$-by-$2$ matrix $\vect{z}(t)=[\vect{x},\vect{y}]=[z_1(t)^\top,\dots, z_n(t)^\top]^\top$.

We propose an example to clarify the concepts of action and opinion, although the model has a broad range of applications. In the context of spelling conventions in the English language, $x_i(t)=+1$ and $x_i(t)=-1$ may represent individual $i$ using the spelling ``center'' and ``centre'', respectively~\cite{Amato2018}. Then, $y_i(t)$ represents individual $i$'s attitude, with a positive and negative $y_i(t)$ representing individual $i$ preferring ``center'' and ``centre'', respectively; the magnitude of $y_i(t)$ indicates the strength of preference. This example clearly illustrates a scenario in which the action taken by an individual may strongly differ from their preference, due to social influence and pressure to conform expected norms. 

We consider a setting in which individuals interact on a two-layer network $\mathcal{G}=(\mathcal V,\mathcal E_A,\mat{A},\mathcal E_W, \mat{W})$, which we assume to be connected, and in which the two nonnegative weight matrices are row-stochastic. Specifically, individuals observe others' actions through an \emph{influence layer} $\mathcal{G}_A=(\mathcal V,\mathcal E_A,\mat A)$ and share opinions across a \emph{communication layer} $\mathcal{G}_W=(\mathcal V,\mathcal E_W,\mat W)$, as illustrated in Fig.~\ref{fig:network}.  

\subsection{Payoff Function}
In our coevolutionary model, we combine the payoff functions of a network coordination game from \eqref{eq:f_function_action} and the Friedkin-Johnsen model from \eqref{eq:f_function_opinion} in a single payoff function. Hence, given state $\vect{z}=(\vect{x},\vect{y})$, we propose the payoff function $f_i(\vect{\zeta},\vect{z})$ for individual $i$ for selecting strategy $\vect{\zeta}=(\zeta_a,\zeta_o)$, with action $\zeta_a\in\{-1,+1\}$ and opinion $\zeta_o \in [-1,1]$:
\begin{align}\label{eq:f_function}
f_i&(\vect{\zeta},\vect{z})=\frac{\lambda_i(1-\beta_i)}{4}\sum_{ j\in \mathcal{V}}a_{ij}\bigg[(1-x_j)(1-\zeta_a)\nonumber
\\&\qquad\qquad\quad+(1+\alpha)(1+x_j)(1+\zeta_a)\bigg]\nonumber\\
&-\frac12\beta_i(1-\lambda_i)(1-\gamma_i)\sum_{j\in\mathcal{V}}w_{ij}(\zeta_o-y_j)^2\nonumber
\\&-\frac12\beta_i(1-\lambda_i)\gamma_i(\zeta_o-u_i)^2-\frac12 \lambda_i\beta_i(\zeta_a-\zeta_o)^2
\end{align}
where $u_i, \gamma_i, a_{ij}, w_{ij}$ are defined in Sections~\ref{sec:coord_game} and \ref{sec:FJ}, and recalled in Table~\ref{tab:parameters}.

\begin{figure}
    \centering
    \includegraphics[width=\columnwidth]{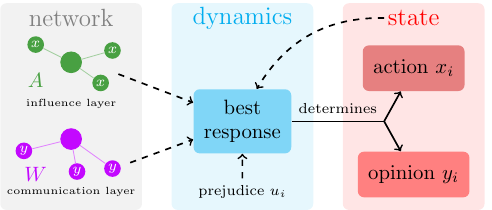}
    \caption{Schematic of the coevolutionary model. }
    \label{fig:schematic}
\end{figure}

The payoff function is made of four terms, as illustrated in Fig.~\ref{fig:schematic}. The first term comes from network coordination games in \eqref{eq:f_function_action}, as discussed in detail in Section~\ref{sec:coord_game}. The second and third terms correspond to the opinion dynamics model in \eqref{eq:f_function_opinion}, discussed in detail in Section~\ref{sec:FJ}. The fourth term is proportional to the squared difference between an individual's action and their opinion, which accounts for an individual's desire to resolve inconsistencies between their actions and opinions. This term, and the weighting parameters $\lambda_i \in [0,1]$ and $\beta_i \in [0, 1]$, underline our proposed method for coupling the evolution of the decision-making and opinion dynamics processes. One will see this clearly in the sequel, when we present the best-response dynamics with respect to the payoff function in \eqref{eq:f_function}. The four terms are weighted by two parameters, $\lambda_i$ and $\beta_i$, representing the contribution of the actions and opinions in determining individual $i$'s payoff, respectively. Specifically, every term of the payoff in \eqref{eq:f_function} that contains action states (the first and the fourth) is multiplied by $\lambda_i$, while those terms that do not contain them are multiplied by $1-\lambda_i$; this is similarly implemented with the parameter $\beta_i$ for opinion states. All the model parameters are summarized in Table~\ref{tab:parameters}.

\begin{table}    \caption{Models variables and parameters. }
    \label{tab:parameters}
    \centering
\begin{tabular}{r|l}
$x_i(t)\in\{-1,+1\}$&action of individual $i$ at time $t$\\
$y_i(t)\in[-1,+1]$&opinion of individual $i$ at time $t$\\
$a_{ij}\in[0,1]$&influence of $j$'s action on $i$\\
$w_{ij}\in[0,1]$&influence of $j$'s opinion on $i$\\
$\alpha\geq 0$&advantage of action $+1$\\
$\lambda_i\in[0,1]$&weight of actions\\
$\beta_i\in[0,1]$&weight of opinions\\
$u_i\in[-1,+1]$&prejudice of individual $i$\\
$\gamma_i\in[0,1]$&attachment to the prejudice of individual $i$
    \end{tabular}\vspace{-10pt}
\end{table}

\begin{remark}
When $\beta_i=0$, then the payoff function in \eqref{eq:f_function} reduces to the payoff function of a coordination game, previously discussed in~\eqref{eq:f_function_action}; for $\lambda_i=0$, it reduces to the one of the Friedkin-Johnsen model, previously discussed in~\eqref{eq:f_function_opinion}. Finally, when $\lambda_i=\beta_i=1$, individual $i$ is totally self-centered and closed to influence from other individuals.
\end{remark}

In the sequel, we explore the payoff function $f_i$ and how it is used to define the coevolutionary model dynamics. For notational convenience, we will sometimes omit explicitly writing the state of the system $\vect{z}=(\vect{x},\vect{y})$ when there is no risk of confusion, and thus we use $f_i(\vect{\zeta})$ to denote $f_i(\vect{\zeta},\vect{z})$. 

Before presenting the update rule of the coevolutionary dynamics, we will present a result on the maximizers of the payoff function in \eqref{eq:f_function}, which allows us to characterize the best-response strategies of the coevolutionary game $\gamma$. Given a state of the system $\vect{z}$ and an individual $i\in\mathcal{V}$, let us define the discriminant quantity $\delta_i(\vect{z})$ as:
\begin{align}\label{eq:delta}
\delta_i(\vect{z})=\;&2\lambda_i\beta_i(1-\lambda_i)\Big[(1-\gamma_i)\sum_{j\in \mathcal{V}}w_{ij}y_j+\gamma_iu_i\Big]\nonumber\\&
+\frac{(1-\beta_i)\lambda_i}{2}\sum_{j\in\mathcal{V}} a_{ij}\Big[2x_j+\alpha(1+x_j)\Big].
\end{align}
The following two results characterize the best-response strategies for the payoff function in \eqref{eq:f_function}. For the sake of readability, the proofs are reported in Appendix~\ref{app:proof_Payoff_Prop}.
\smallskip

\begin{proposition}\label{prop:payoff}
For the coevolutionary game $\gamma=(\mathcal{V},\mathcal{A},\vect{f})$ on the two-layer network $\mathcal{G}$, with the $i$th entry of $\vect f$ defined in \eqref{eq:f_function}, and the function $\delta_i(\vect{z})$ defined in \eqref{eq:delta}, consider a player $i\in\mathcal V$. Suppose also $\lambda_i \in[0,1]$ and $\beta_i \in(0,1]$. If $\delta_i(\vect{z})\neq 0$, then the best-response strategy $\mathcal{B}_i\big(f_i(\cdot,\vect{z}(t))\big)=(\zeta_a,\zeta_o)$, with
\begin{subequations}
\begin{align}
\zeta_a&=\text{sgn}{(\delta_i)}\\
\zeta_o&=(1\!-\!\lambda_i)\Big[(1\!-\!\gamma_i)\sum_{j\in\mathcal{V}}w_{ij}y_j\!+\!\gamma_iu_i\Big]\!+\!\lambda_i\text{sgn}(\delta_i)
\end{align}
\end{subequations}
If $\delta_i(\vect{z})=0$, then $\mathcal{B}_i\big(f_i(\cdot,\vect{z}(t))\big)=\{(+1,\zeta_o),(-1,\bar \zeta_o)\}$, with
\begin{subequations}
\begin{align}
\zeta_o& =(1-\lambda_i)\big[(1-\gamma_i)\sum\nolimits_{j\in\mathcal{V}}w_{ij}y_j+\gamma_iu_i\big]+\lambda_i \\
\bar \zeta_o& =(1-\lambda_i)\big[(1-\gamma_i)\sum\nolimits_{j\in\mathcal{V}}w_{ij}y_j+\gamma_iu_i\big]-\lambda_i.
\end{align}
\end{subequations}
\end{proposition}
\smallskip

Note that Proposition~\ref{prop:payoff} illustrates how the coupling between opinions and actions due to the fourth term in \eqref{eq:f_function} yields a nontrivial best-response strategy when the action and opinion are jointly revised. Importantly, the update rules for $x_i(t)$ and $y_i(t)$ cannot be decoupled, as they are intertwined by the quantity $\delta_i(\vect{z})$, which depends both on opinions and actions. We also point out that, although the best-response expression in Proposition~\ref{prop:payoff} looks complex, the expression comes from two separate processes well established in the literature (network coordination games and Friedkin-Johnsen opinion dynamics) being coupled together by a term that, intuitively, suggests an individual has a desire to resolve discrepancies between their action and opinion. 

\begin{remark}\label{rem:best}
In general, the  best-response strategy might  not be unique. A trivial scenario is the limit case $\beta_i=0$, in which opinions do not influence the payoff and \eqref{eq:f_function} reduces to \eqref{eq:f_function_action}, and every $\zeta_o\in[-1,1]$ provides the same payoff. More interestingly, for $\beta_i\neq 0$, the payoff function $f_i$ has two maximizers when $\delta_i=0$: one with positive action $\zeta_a=+1$ and one with negative action $\zeta_a=-1$. In the sequel, we make a simple assumption on how the individuals resolve the issue of multiple maximizers.
\end{remark}

\subsection{Dynamics}\label{ssec:dynamics}

We are now in position to define the update dynamics for the individuals' actions and opinions. We consider an asynchronous rule. Specifically, at each discrete time instant $t\in \mathbb Z_+$, a single\footnote{Our model can be readily extended to consider a nonsingleton set of individuals activating at each time instant, or all individuals (corresponding to synchronous updating among the population).} individual $i \in \mathcal{V}$ becomes active and updates their opinion and action simultaneously. In the next section when we present the main convergence result of this paper, we provide an explicit assumption on how individuals are selected to activate at each time step. However, our model is general to allow convergence analysis of $\vect z(t)$ under different assumptions on agent activation sequences, and so no explicit assumption is imposed in this section, where we simply focus on presentation of the individual dynamics. 

If individual $i \in \mathcal{V}$ becomes active at time $t$, then they revises their state $\vect{z}_i(t)$ according to a best-response update, that is, 
\begin{equation}\label{eq:br_xy}
    \vect{z}_i(t+1) \in \mathcal{B}_i\big(f_i(\cdot,\vect{z}(t))\big)
\end{equation}
where $\mathcal{B}_i$ is the best-response strategies defined in \eqref{eq:best_response_generic} and $f_i$ is the payoff in \eqref{eq:f_function}. As we have observed in Remark~\ref{rem:best}, in general, the best response might not be  unique, and in such scenarios, we should select a tie-break rule to unequivocally determine the system evolution. Proposition~\ref{prop:payoff} gives explicit expressions for the best-response strategies for the payoff function $f_i$, which allows us to write explicit expressions for \eqref{eq:br_xy}, and resolve ties (which occur if $\delta_i=0$).

In particular, we make explicit the best-response update rule of the coevolutionary dynamics as follows. For active individual $i$ with $\beta_i\in(0,1]$, the state is updated as
\begin{align}\label{eq:xUpdate}
&x_i(t+1)=\mathcal{S}(\delta_i(\vect{z}(t))\\\label{eq:yUpdate}
&y_i(t+1)=(1-\lambda_i)\Big[(1-\gamma_i)\sum_{j\in\mathcal{V}}w_{ij}y_j(t)+\gamma_iu_i\Big]+\lambda_i\mathcal{S}(\delta_i)
\end{align}
where
\begin{equation} \label{eq:S_delta}
\mathcal{S}(\delta_i(\vect{z}(t)))=\begin{cases} +1 & \text{if }\delta_i(\vect{z}(t))>0\\-1 &  \text{if }\delta_i(\vect{z}(t))<0\\x_i(t) &  \text{if }\delta_i(\vect{z}(t))=0.\end{cases}
\end{equation} 
The above update rule resolves the tie in which both actions $-1$ and $+1$ give the same payoff. Specifically, we have adopted a conservative method for resolving ties, enforcing an agent to stick with the current action when both possible actions give equal reward, consistent with the social psychology literature on the presence of inertia in decision-making and status-quo bias~\cite{samuelson1988statusquo}. Finally, if $\beta_i = 0$, following a similar approach, we set that $y_i(t+1) = y_i(t)$, which translates to $y_i(t) = y_i(0)$ for all $t\geq 0$, while $x_i(t+1)$ is given in \eqref{eq:xUpdate}. 

To summarize, the proposed coevolutionary model is formulated as a best-response dynamics for the \emph{coevolutionary game} $\gamma=(\mathcal{V},\mathcal{A},\vect{f})$ on a two-layer network $\mathcal{G}$, with a two-dimensional set of strategies $\mathcal{A}=\{-1,+1\}\times[-1,+1]$, and the payoff vector is defined with its $i$th entry in \eqref{eq:f_function}.

\section{Convergence Analysis}\label{sec:analysis}

In this section, we establish a general convergence result for the proposed coevolutionary dynamics for homogeneous agents, which is one of the main theoretical contributions of this work. To keep the exposition clear, the proofs for each theorem, lemma or corollary are presented in the Appendix.    \smallskip

In Section~\ref{sec:model}, we presented our coevolutionary model in a game-theoretic framework, casting the update mechanism as a best-response dynamics to the payoff function \eqref{eq:f_function}. Our convergence analysis is thus inspired by game-theoretic methods, in particular the theory of potential games~\cite{monderer1996potential}. To begin, we introduce the following function and establish two associated technical results that prove the function is in fact a potential function for the coevolutionary game:
\begin{align}\label{eq:phi}
\Phi&(\vect{z})=\sum_{i\in\mathcal{V}}\sum_{j\in\mathcal{V}\setminus\{i\}} \eta_i \frac{a_{ij}}{2}\bigg[(1+\alpha)(1+x_j)(1+x_i)\nonumber\\&\;\;\qquad+(1-x_i)(1-x_j)\bigg]\nonumber\\
&-\frac{1}{2}\sum_{i\in\mathcal{V}}\sum_{j\in\mathcal{V}}\frac{w_{ij}}{2}(y_i-y_j)^2
-\frac{1}{2}\sum_{k\in\mathcal{V}}\frac{\gamma_i}{(1-\gamma_i)}(y_k-u_k)^2
\nonumber\\&-\frac12\sum_{k\in\mathcal{V}}\frac{\lambda_i}{(1-\lambda_i)(1-\gamma_i)}(y_k-x_k)^2
\end{align}
where $\eta_i={\lambda_i(1-\beta_i)}/{4\beta_i(1-\lambda_i)(1-\gamma_i)}$.

The following lemma, whose proof is reported in Appendix~\ref{app:Lemma1}, connects the potential function to the payoff function in \eqref{eq:f_function} under an assumption on the parameters.\smallskip

\begin{assumption}\label{as:uniform}
Suppose that all the parameters are homogeneous, i.e., $\lambda_i=\lambda\in(0,1)$, $\gamma_i=\gamma\in(0,1)$, and $\beta_i=\beta\in(0,1)$ for all $i\in\mathcal{V}$. Suppose further that the two layers are undirected, that is, $\mat{W}=\mat{W}^\top$, and $\mat{A}=\mat{A}^\top$, and that each individual has a self-loop on the influence layer, that is, $a_{ii}>0$ for all $i\in\mathcal{V}$.
\end{assumption}\smallskip

\begin{lemma}\label{lem:potentialGame}
Under Assumption~\ref{as:uniform}, the coevolutionary game $\gamma=(\mathcal{V},\mathcal{A},\vect{f})$ on the two-layer network $\mathcal{G}$ is a generalized ordinal potential game with potential function $\Phi$, defined in \eqref{eq:phi}.
\end{lemma}\smallskip

Next, we provide an explicit lower bound on the increase in the potential function if an agent that activates changes action according to the coevolutionary dynamics in Eqs.~(\ref{eq:xUpdate}) and (\ref{eq:yUpdate}). Its proof is reported in Appendix~\ref{app:Lemma1}\smallskip

\begin{cor}\label{cor:potential_LB}
Consider the coevolutionary model on the two-layer network $\mathcal{G}$ in which individuals update their  actions and opinions according to \eqref{eq:xUpdate} and \eqref{eq:yUpdate}, respectively. Suppose that Assumption~\ref{as:uniform} holds. If agent $i\in \mathcal{V}$ is active at time $t$, and $x_i(t) \neq x_i(t+1)$, then
\begin{equation}\label{eq:PhiDifPositive}
    \Phi(\vect{z}(t+1))-\Phi(\vect{z}(t)) > \min_{k\in\mathcal V}\xi\lambda(1-\beta)a_{kk} > 0\end{equation} 
where $\xi = [\beta(1-\lambda)(1-\gamma)]^{-1} > 0$.
\end{cor}\smallskip

In Section~\ref{ssec:dynamics}, we detailed the general model dynamics, whereby a single agent $i\in \mathcal V$ activates at each discrete time $t\in \mathbb Z_+$ to update their state $\vect{z_i}(t)=(x_i(t), y_i(t))$. However, there can be a number of different mechanisms governing the sequence of agent activations. In our work, we secure a general convergence result by assuming a particular mechanism, as we now detail.

\begin{assumption}\label{as:activation}
There exists a $T \in \mathbb{Z}_+$ such that in every time window $[t; t + T), \; \forall t \in \mathbb Z_+$, each agent $i \in\mathcal{V}$ activates at least once\footnote{Note this implies $T \geq n$.}.
\end{assumption}\smallskip

This assumption ensures that every agent will activate at least once in $T$ time steps. It covers a broad range of scenarios, since $T$ can be much greater than $n$ (allowing agents to update multiple times inside of $T$ time steps), and imposes no restriction or requirement on the ordering of the agent activations. Other assumptions on the activation sequence can be considered, which may lead to different convergence results. For instance, one could assume that at every time step, every agent has a positive probability of becoming active, or that every agent activates infinitely often as $t\to\infty$~\cite{riehl2018survey}. 

Although we proved in Lemma~\ref{lem:potentialGame} that the coevolutionary game is a generalized ordinal potential game, we cannot exploit existing convergence results on potential games such as those detailed in~\cite{monderer1996potential}. This is because in our game, the strategy space for each player (individual) is a Cartesian product of a finite strategy set (actions) and a continuous strategy set (opinions), and as noted below Proposition~\ref{prop:payoff}, the action and opinion updating dynamics cannot be decoupled. Significant additional analysis is therefore required. We first present a convergence result for the opinion vector under the assumption that the action vector is fixed, which is key for the proof of our main convergence result, reported in the following. For the sake of readability, the proof is postponed to Appendix~\ref{app:Lemma2}. \smallskip

\begin{lemma}\label{lem:opinion_convergence}
Consider a modified coevolutionary model on the two-layer network $\mathcal{G}$, where individuals update their opinions according to \eqref{eq:yUpdate} while the actions are held constant, $x_i(t+1) = x_i(t) = x_i^*$ for all $t\geq 0$, and let action vector be denoted by $\vect x^* \in \{-1, +1\}^n$, Then, under Assumption~\ref{as:uniform}, there is a unique opinion vector $\vect y^* \in [-1, 1]^n$ that is the unique equilibrium of the opinion dynamics in \eqref{eq:yUpdate}. Further, if Assumption~\ref{as:activation} also holds, then $\lim_{t\to\infty} y_i(t) = y_i^*$ asymptotically. 
\end{lemma}\smallskip

Finally, we are ready to present the main convergence result of our paper, whose proof is reported in Appendix~\ref{app:Theorem1}. This result establishes that for all initial conditions, the system converges to a steady state.\smallskip

\begin{theorem}\label{th:convergence}
Consider a coevolutionary model on the two-layer network $\mathcal{G}$ in which individuals update their opinions and actions according to \eqref{eq:yUpdate} and \eqref{eq:xUpdate}, respectively. Suppose that Assumptions~\ref{as:uniform} and \ref{as:activation} hold.  Then, for all initial conditions $\vect z(0) \in \{-1,+1\}^n \times [-1,1]^n$, the state of the system $\vect{z}(t)$ converges to a steady state $\vect{z}^*$. Specifically, the action vector $\vect{x}(t)$ converges to $\vect{x^*}$ in finite time, while the opinion vector $\vect{y}(t)$ converges to $\vect{y^*}$ asymptotically.
\end{theorem}\smallskip

Thus, the key conclusion on the dynamics of the coevolutionary model is that under Assumption~\ref{as:uniform} on the parameters, and Assumption~\ref{as:activation} on the activation mechanism for the agents, the actions and opinions of all individuals converge to an equilibrium $\vect z^* = (\vect x^*, \vect y^*)$, for all initial conditions. Moreover, $\vect z^*$ is a Nash equilibrium of the coevolutionary game. However, there are generally speaking a number of different Nash equilibria which the system can converge to, and it is difficult to provide insight into regions of attraction for the equilibria due to the highly nonlinear dynamics (although Lemma~\ref{lem:opinion_convergence} indicates that for a given action equilibrium $\vect x^*$, the opinion equilibrium $\vect y^*$ is unique). In fact, Assumption~\ref{as:activation} places little restriction on the agent activation sequence, and thus for the same $\vect z(0)$, it is possible to converge to two different Nash equilibria $\hat{\vect z}^*$ and $\bar{\vect z}^*$, for different agent activation sequences. In the next section, we provide a region of attraction for a specific type of equilibrium, under further conditions.

Before concluding this section, we would like to stress that the fundamental properties detailed in Lemma~\ref{lem:potentialGame} and Corollary~\ref{cor:potential_LB}, which are key for the convergence analysis, are structural properties of the coevolutionary dynamics and are independent of the activation mechanism. This suggests that future work may consider convergence analysis under different assumptions on the activation mechanism.

\section{Emergent Real-world Phenomena}\label{sec:phenomena}

In the rest of this paper, we will utilize the proposed coevolutionary model and the convergence result in Theorem~\ref{th:convergence} to investigate some specific emergent phenomena of the model, which may have several interesting real-world applications. Besides the phenomenon of unpopular norms and shifts in social conventions, which could be reproduced and studied following the paradigm proposed for a similar coevolutionary model in~\cite{zino2020coevolutionary,zino2020two}, we will show that our model can reproduce polarization and pluralistic ignorance. Specifically, in Section~\ref{sec:polarization}, we derive analytical results, supported by numerical simulations, to show that the proposed coevolutionary dynamics can lead to the emergence and persistence of polarization in a population, which is a typical phenomenon of many social systems~\cite{Fiorina2008,Cinelli2021}. Then, in Section~\ref{sec:pluralistic}, we will numerically illustrate how the phenomenon of pluralistic ignorance, which is well-known and extensively studied in the social psychology literature~\cite{merton1968social_book,prentice1993pluralistic}, can emerge in a population over time.

\subsection{Polarization}\label{sec:polarization}

Here, we investigate how the model may lead to the emergence of stable polarized states, in which a portion of the population opts for and supports one action, and the rest of the population takes and supports the other one. Given the set of agents $\mathcal{V}$, we denote by $(\mathcal{V}_p, \mathcal{V}_n)$ a partitioning of the agents into two disjoint sets $\mathcal{V}_p$ and $\mathcal{V}_n$ satisfying $\mathcal{V}_p\bigcap\mathcal{V}_n=\emptyset$, $\mathcal{V}_p\bigcup\mathcal{V}_n=\mathcal{V}$ and $\mathcal{V}_p \neq \emptyset$ and $\mathcal{V}_n \neq \emptyset$. With this notion of a partition in place, we introduce the following definitions of polarization in the network.

\begin{definition}\label{def:polarized_state}
Given a state of the system $\vect z=(\vect{x},\vect{y})\in\{-1,1\}^n\times [-1,1]^n$, we say that
\begin{itemize}
    \item the action vector $\vect x$ is \emph{polarized} if there exists a partition of the agents ($\mathcal{V}_p$, $\mathcal{V}_n$) such that $x_i=-1$, for all $i\in\mathcal{V}_n$ and $x_j=+1$, for all $j\in\mathcal{V}_p$.
    \item the opinion vector $\vect y$ is \emph{polarized} if there exists a partition of the agents ($\mathcal{V}_p$, $\mathcal{V}_n$) such that $y_i<0$, for all $i\in\mathcal{V}_n$ and $y_j>0$, for all $j\in\mathcal{V}_p$.
    \item the state $\vect{z}$ is \emph{polarized} if there exists a partition of the agents ($\mathcal{V}_p$, $\mathcal{V}_n$) such that $x_i=-1$ and $y_i<0$, for all $i\in\mathcal{V}_n$, and $x_j=1$ and $y_j>0$, for all $j\in\mathcal{V}_p$.
\item the state $\vect{z^*}$ is a \emph{polarized equilibrium} if it is a polarized state, and it is an equilibrium of the coevolutionary model in Eqs.~(\ref{eq:yUpdate})--(\ref{eq:xUpdate}).
\end{itemize}
\end{definition}

Briefly, a state is polarized if the population is split into two clusters, where individuals in the same cluster have the same action, but different actions between different clusters. While the opinions of individuals within the same cluster have the \textit{same sign}, the opinions are not necessarily of the \textit{same magnitude} (and $y_i$ does not have to be equal to $+1$ or $-1$).

In the following, we study the polarization problem under the simplifying assumptions that i) the communication layer and the influence layer coincide, ii) no action has a payoff advantage, and iii) agents are not attached to any existing prejudice. These assumptions are gathered in the following.\smallskip

\begin{assumption}\label{as:polarization}
We assume that $\mat A=\mat W$, $\alpha=0$, and $\gamma = 0$.
\end{assumption}\smallskip

Under Assumption~\ref{as:polarization}, the opinion and action updates rules are simplified as follows \begin{equation} \label{eq:yxUpdate_POL}
\left\{\begin{aligned}
y_i(t+1)&=(1-\lambda)\sum\nolimits_{j\in\mathcal V}w_{ij}y_j(t)+\lambda\mathcal{S}(\delta_i(\vect{z}(t))
\\x_i(t+1)&=\mathcal{S}(\delta_i(\vect{z}(t)))
\end{aligned}\right.
\end{equation} 
where
\begin{equation} \label{eq:deltaPol}
\delta_i(\vect{z})=2\lambda\beta(1-\lambda)\sum_{j\in \mathcal{V}}w_{ij}y_j+(1-\beta)\lambda\sum_{j\in\mathcal{V}} w_{ij}x_j.
\end{equation} 

The following result provides a sufficient condition for the existence of a polarized equilibrium, with the proof in Appendix~\ref{app:polarized}.\smallskip

\begin{theorem}\label{th:polarized}
Consider the coevolutionary model on the two-layer network $\mathcal{G}$ under Assumptions~\ref{as:uniform}--\ref{as:polarization}, in which individuals update their opinions and actions according to \eqref{eq:yxUpdate_POL}. Suppose there exists a partitioning  of the agents ($\mathcal{V}_p$,$\mathcal{V}_n$) with 
\begin{equation} \begin{aligned}
\underline d_p &\coloneqq \min_{i\in \mathcal{V}_p}\sum\nolimits_{j\in\mathcal{V}_p}w_{ij}\quad \bar d_p \coloneqq \max_{i\in \mathcal{V}_p}\sum\nolimits_{j\in\mathcal{V}_p}w_{ij}\\
\underline d_n &\coloneqq \min_{i\in \mathcal{V}_n}\sum\nolimits_{j\in\mathcal{V}_n}w_{ij}\quad \bar d_n \coloneqq \max_{i\in \mathcal{V}_n}\sum\nolimits_{j\in\mathcal{V}_n}w_{ij}
\end{aligned}
\end{equation} 
such that the following three conditions are verified:
\begin{equation} \label{eq:pol_y_cond}
\lambda>\max\bigg(\frac{\bar d_p-\underline d_n}{1+\bar d_p-\underline d_n},\frac{\bar d_n-\underline d_p}{1+\bar d_n-\underline d_p}\bigg)
\end{equation} 
\begin{equation} \label{eq:pol_xp_cond}
    \frac{\lambda (\underline{d}_p + \bar{d}_n - 1) + \underline{d}_p - \bar{d}_n}{1-(1-\lambda)(\underline d_p+\bar d_n-1)}+ \frac{(1-\beta)(2\underline d_p-1)}{2 \beta  (1-\lambda )} > 0
\end{equation} 
and
\begin{equation} \label{eq:pol_xn_cond}
    \frac{\lambda (\underline{d}_n + \bar{d}_p - 1) + \underline{d}_n - \bar{d}_p}{1-(1-\lambda)(\underline d_n+\bar d_p-1)}+\frac{(1-\beta)(2\underline d_n-1)}{2 \beta  (1-\lambda )} >0.
\end{equation} 
Then the coevolutionary model in~\eqref{eq:yxUpdate_POL} has a polarized equilibrium $\vect{z}^*$ with respect to the partitioning $(\mathcal{V}_p,\mathcal{V}_n)$. 
\end{theorem}\smallskip

Theorem~\ref{th:polarized} indicates that a polarized equilibrium $\vect z^*$ exists with respect to a partitioning $(\mathcal{V}_p, \mathcal{V}_n)$ if the three conditions of Eqs.~(\ref{eq:pol_y_cond})--(\ref{eq:pol_xn_cond}) are satisfied. While the latter two conditions are not easy to digest, all three conditions can be easily computed. However, the result provides no indication of the basin of attraction for $\vect z^*$. Next, we provide a more conservative sufficient condition, which however allows to prove not only the existence of a polarized equilibrium but also a notion a region of attraction from a set of initial conditions. The proof is reported in Appendix~\ref{app:polarized_convergence}. \smallskip

\begin{theorem}\label{th:polarized_convergence}
Consider a coevolutionary model on the two-layer network $\mathcal{G}$ under Assumptions~\ref{as:uniform}--\ref{as:polarization}, in which individuals update their opinions and actions according to \eqref{eq:yxUpdate_POL}. Assume that there exists  a partitioning of the agents ($\mathcal{V}_p$, $\mathcal{V}_n$) such that
\begin{equation} \label{eq:condition_polarized}\begin{cases}
\sum_{j\in\mathcal{V}_p}w_{ij}> \max\left(\frac{1-2\lambda}{1-\lambda},\frac12\Big(1+\frac{\beta(1-\lambda)}{1-\beta\lambda}\Big)\right),  &\text{for } i\in\mathcal{V}_p
\\
\sum_{j\in\mathcal{V}_n}w_{ij}> \max\left(\frac{1-2\lambda}{1-\lambda},\frac12\Big(1+\frac{\beta(1-\lambda)}{1-\beta\lambda}\Big)\right),  &\text{for }i\in\mathcal{V}_n.
\end{cases}\end{equation} 
Then, if the initial condition $\vect{z}(0)$ is a polarized state, $\vect{z}(t)$ will be polarized for any $t\in\mathbb Z_+$, and $\lim_{t\to\infty} \vect{z}(t) = \vect{z}^*$, where $\vect z^*$ is the polarized equilibrium with respect to the partition $(\mathcal{V}_p,\mathcal{V}_n)$.
\end{theorem}\smallskip

Theorem~\ref{th:polarized_convergence} provides easier to check conditions, viz. \eqref{eq:condition_polarized}, for the existence of a polarized equilibrium $\vect z^*$. Moreover, under these conditions, a large region of attraction for $\vect z^*$ is identified, being the set of initial conditions $\vect z(0)$ which is polarized with respect to the partition $(\mathcal{V}_p,\mathcal{V}_n)$. From another perspective, the set of polarized states with respect to $(\mathcal{V}_p,\mathcal{V}_n)$ is \textit{a positive invariant set} of the system dynamics. \eqref{eq:condition_polarized} provides insight into the role of the model parameters on the emergence of polarized equilibria. In fact, we observe that the conditions become monotonically stricter as $\lambda$ decreases and $\beta$ increases. This seems to suggest that polarization emerges more easily in scenarios in which individuals give higher weight to observed actions rather than shared opinions. 

This result demonstrates that polarization can emerge and become stable in a population where opinions and actions coevolve. Such a finding is notable, since in the opinion dynamics literature, polarization has been typically attributed to antagonistic interactions under restrictive signed network structure assumptions~\cite{altafini2012consensus,xia2015structural_opinions} or strong biased assimilation~\cite{dandekar2013biased_degroot,xia2020bias}. Here, we show that when an individual's opinion coevolves with their action, polarization can emerge purely through coordination and social influence, similar to observations from other models with opinions and quantized actions~\cite{Martins2008coda,Chowdhury2016coda,Ceragioli2018quantized}. In the following example, we show via simulation results that the behavior proved in Theorem~\ref{th:polarized_convergence} seems to be resistant to small perturbations of the network and that the region of attraction of polarized equilibria may be larger, whereby polarization can emerge even from nonpolarized initial states.

\begin{example}\label{ex:polarization}
We consider  $20$ interacting individuals divided into two groups $\mathcal{V}_p=\{1,\ldots,10\}$ and $\mathcal{V}_n=\{11,\ldots,20\}$. Groups are generated as directed Erdős–Rényi random graphs~\cite{Boccaletti2014}, where each link is present with probability $p=0.2$, independent of the other edges. Then, the two groups are connected by two links between two pairs of nodes (one from each group), selected uniformly at random. Last, weights are selected randomly from a uniform distribution and rescaled such that the overall influence matrix is row-stochastic. The network obtained is presented in Fig.~\ref{fig:pol_G}.

We set $\alpha=\gamma=0$, $\beta=0.5$, and $\lambda=0.6$. With these parameters, the inequality conditions in Theorem~\ref{th:polarized_convergence} are satisfied. However, Assumption~\ref{as:uniform}  is not satisfied, since self-loops are not present for all the nodes. Furthermore, we initialize the system with $\vect z(0)$, where most of the nodes in $\mathcal{V}_p$ have $x_i(0)=\text{sgn}(y_i(0))=+1$ and  most of the nodes in $\mathcal{V}_n$ have $x_i(0)=\text{sgn}(y_i(0))=-1$. However, we randomly assign some exceptions to this rule, so that the initial state $\vect z(0)$ is not polarized, as shown in Fig.~\ref{fig:pol_G0}. We simulate the system by selecting one agent uniformly at random at each time step to activate and update their states, checking that all agents activate at least once during the simulation.

While the initial state $\vect z(0)$ is outside the region of attraction identified in Theorem~\ref{th:polarized_convergence}, the numerical simulation reported in Fig.~\ref{fig:pol_G} shows that the coevolutionary model still converges to the polarized equilibrium with respect to $(\mathcal{V}_p, \mathcal{V}_n)$. This suggests the region of attraction for polarized equilibria may be larger than that identified in our theoretical findings.  

\begin{figure}
    \centering
    \begin{subfigure}[b]{0.49\columnwidth}
     \centering\includegraphics{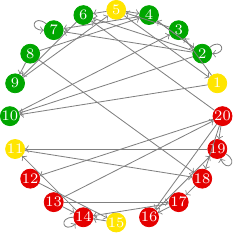}   
    \caption{Initial actions. }
    \label{fig:pol_G0}
     \end{subfigure}
    \begin{subfigure}[b]{0.49\columnwidth}
     \centering\includegraphics{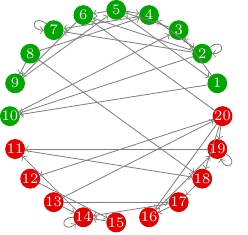}
    \caption{Final actions.}
    \label{fig:pol_Gf}
     \end{subfigure}\\
         \begin{subfigure}[b]{\columnwidth}
     \centering\includegraphics{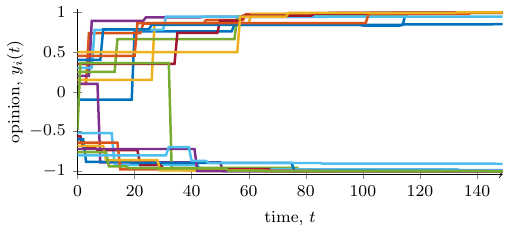}   
    \caption{Temporal evolution of the opinions.}
    \label{fig:pol_y}
     \end{subfigure}
     \caption{Network and simulation of Example~\ref{ex:polarization}. In (a) and (b), the network and the actions are shown at time $t=0$ and $t=150$, respectively. Nodes are colored in green if $\text{sgn}(y_i)=x_i=+1$, in red if $\text{sgn}(y_i)=x_i=-1$, and in yellow if $\text{sgn}(y_i)\neq x_i$. In (c), the temporal evolution of the opinion are shown to converge to a polarized scenario. \vspace{-10pt}}
     \label{fig:pol_G}
\end{figure}
\end{example}

We conclude this section by presenting a special case in which all the agents have a uniform influence at the partition-level, that is, there exists a partition $(\mathcal{V}_p,\mathcal{V}_n)$ such that
\begin{equation}\label{eq:uniform}
    \sum_{j\in\mathcal{V}_p}w_{ij}=d,\,\forall\, i\in\mathcal{V}_p\qquad \sum_{j\in\mathcal{V}_n}w_{ij}=d,\,\forall\, i\in\mathcal{V}_n.
\end{equation} In this scenario, a complete characterization of the polarized equilibrium can be performed, showing the emergence of a \emph{bipartite consensus} state, in which all the agents in the same partition have the same opinion. The proof is in Appendix~\ref{app:bipartite_consensus}. 

\begin{cor}\label{cor:bipartite_consensus}
Consider a coevolutionary model on the two-layer network $\mathcal{G}$ under Assumptions~\ref{as:uniform}--\ref{as:polarization}, in which individuals update their opinions and actions according to \eqref{eq:yxUpdate_POL}. Suppose that there exists a partition $(\mathcal{V}_p,\mathcal{V}_n)$ such that \eqref{eq:uniform} holds true for some $d>1/2$. Then there exists a polarized equilibrium in the form
\begin{equation} \label{eq:uniform1}
y_i^*= y^+\,,\forall\, i\in \mathcal{V}_p \qquad y_j^* = y^-\,,\forall\, j\in \mathcal{V}_n,
\end{equation} 
with
\begin{equation} \label{eq:uniform2}
y^+=\frac{\lambda}{1-(1-\lambda)(2d-1)}\quad y^-=-\frac{\lambda}{1-(1-\lambda)(2d-1)}.
\end{equation} 
\end{cor}

\subsection{Pluralistic Ignorance}\label{sec:pluralistic}

Here, we explore how the social psychological phenomenon of \textit{pluralistic ignorance}~\cite{merton1968social_book} can be observed as an emergent behavior in our model. Pluralistic ignorance refers to a situation whereby a significant number of individuals in a population hold an incorrect assumption about the thoughts, feelings, or opinions of other members. The most common such situation is when a group of individuals underestimates the number of others who hold similar opinions. Such a phenomenon has been widely studied in the literature, with empirical examples concerning overestimation of support among white Americans for racial segregation in 1960s America~\cite{ogorman1975pluralistic}, or underestimation of the rejection of alcohol consumption practices on university campuses in the 1990s~\cite{prentice1993pluralistic}.

In particular, we revisit the classical work of Prentice and Miller~\cite{prentice1993pluralistic}, which reported field studies on the opinions and behaviors of undergraduate students at Princeton University concerning the culture of rampant alcohol consumption. In~\cite{prentice1993pluralistic}, the authors studied how individuals internalized social norms when exposed to pluralistic ignorance over time. They surveyed via telephone a random sample of 25 female and 25 male sophomore students on Princeton campus, first in September at the beginning of the semester year, and then again in December. The subjects were asked several questions about Princeton's alcohol policies, their own opinion toward drinking alcohol, and their estimate of the average student's attitude. The students ranked their own comfort level and their estimate of the average student's comfort level with alcohol drinking, using a number between 0 and 10.

In the following example, we show that with reasonable choices of parameters, our model can qualitatively reproduce the empirical data reported in the study described above.

\begin{example}
In order to apply our model to the work of Prentice and Miller~\cite{prentice1993pluralistic}, we let $\vect y=[\vect y_f^\top, \vect y_m^\top]^\top$, where $\vect y_f,\vect y_m\in[-1,1]^{25}$ are the vectors of attitudes (opinions) of female and male subjects, respectively. In addition, we define $\vect x=[\vect x_f^\top,\vect x_m^\top]^\top$, where $\vect x_f, \vect x_m\in\{-1,1\}^{25}$ represent the vectors of behavior (actions) of female and male students, respectively.
The mean value of $\vect x$, denoted by $ \bar{x}$, is a reasonable choice to represent the ``social norm", as it measures the average behavior of the society.

The initial states are reproduced randomly using a normal distribution with the mean and standard deviation values reported in~\cite[Study~3]{prentice1993pluralistic}. Then, the initial state values are rescaled to fit the state space of our model: $x_i=-1$ and $x_i=+1$ represent the rejection and embracing of heavy alcohol consumption by agent $i$, respectively. Consistent with our model definition, $y_i$ is individual $i$'s preference towards the choices of rejecting or supporting alcohol consumption. In addition, we set $\vect u=\vect y(0)$. 

Prentice and Miller discuss that ``men are simply more inclined to react to feeling deviant from the norm with conformity"~\cite{prentice1993pluralistic}. In our model, this can be reasonably interpreted as $\lambda_m>\lambda_f$, since a larger $\lambda$ means that observed actions have a great impact on the payoff. In context, male individuals strongly favor adopting the action currently in the majority (the social norm), and in their opinion update they place more weight on their own action, which is influenced by the social norm. On the other hand,~\cite{prentice1993pluralistic} also states that ``men might be expected to feel greater pressures to learn to be comfortable with alcohol". In our setting, this can be approached by adjusting $\gamma_m<\gamma_f$. As $\vect u=\vect y(0)$, $\gamma_m<\gamma_f$ implies that male subjects tend to be less attached to their initial opinion.
 
Considering the observations above,  we set $\alpha=0$, $\lambda_f=0.2$, $\lambda_m=0.6$, $\beta_f=\beta_m=0.4$,  $\gamma_f=0.8$, and $ \gamma_m=0.3$. In addition, we generate a two-layer graph with uniformly random weights to represent the student network; both layers are Erdős–Rényi random graphs, with each link present with probability $p$. We assume the influence layer  denser relative to the communication layer by setting $p = 0.1$ and $p = 0.04$, respectively.  This reflects that opinions are mostly private and shared between few friends, while actions are public and observed within the whole campus environment. 
 \begin{figure}
    \centering
     \includegraphics{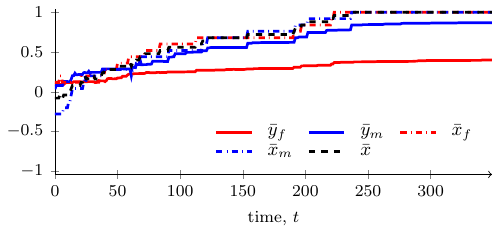}
    \caption{Simulation illustrating the emergence and resolution of pluralistic ignorance among female and male agents, respectively, closely following the findings in~\cite{prentice1993pluralistic}.\vspace{-10pt} }
    \label{fig:pluralistic_ignorance}
\end{figure}

Figure~\ref{fig:pluralistic_ignorance} plots the temporal evolution of the mean values of $\vect y_f$, $\vect y_m$, $\vect x_f$, and $\vect x_m$. According to~\cite{prentice1993pluralistic}, at the start of their experiment in September, it was found that both male and female students displayed strong pluralistic ignorance, whereby the majority of students were privately against the social norm of heavy alcohol consumption, but estimated that their peers preferred heavy alcohol consumption, and all individuals consumed alcohol to a greater degree than they were comfortable with. In our simulation, we initialize individuals'  opinions and  behaviors with the same discrepancy.

Prentice and Miller reported that by December when they interviewed the participants again, male participants had tended to resolve pluralistic ignorance over time by internalization of the social norm, moving their private attitudes to conform to strong drinking culture. In contrast, female participants reacted with alienation by mostly retaining their private beliefs, but continuing to participate in the social norm of heavy alcohol consumption~\cite{prentice1993pluralistic}.

Our simulation is qualitatively similar to the outcomes described by Prentice and Miller. In the simulation, the male agents tend to internalize the social norm, $\bar x$, by moving their  opinions, viz. $\bar y_m$ to be closer to the social norm. On the other hand, female agents remain attached to their initial opinion, and their opinion, viz. $\bar y_f$ remains significantly less supportive of heavy alcohol consumption.  Additional simulations, omitted due to space constraints, confirm that our findings are robust to small changes to the values of the model parameters, and even heterogeneity among parameters. 
\end{example}

\section{Conclusions}\label{sec:conclusions}
In this paper, we proposed a novel coevolutionary model of actions and opinions that captures the complex interplay between opinion formation and decision-making processes in a social network, utilizing a game-theoretic framework. We provided a rigorous analysis of the model, establishing finite-time convergence of the actions and asymptotic convergence for the opinions, for any initial condition. Then, we utilized the coevolutionary model to explore various real-world phenomena. Besides the emergence of unpopular norms, which were studied on a simplified version of the model~\cite{zino2020coevolutionary,zino2020two}, the proposed model can represent stable polarization and the emergence of pluralistic ignorance.

The vast range of real-world phenomena that our model can represent and its amenability to perform rigorous analytical studies pave the way for a wide horizon of future research. First, our convergence result is limited to homogeneous scenarios, and it should be generalized to heterogeneous populations, as a key step toward studying real-world complex social systems which may have high degrees of heterogeneity~\cite{Boccaletti2014}. Second,  our simulations suggest polarization can emerge due to the coordination and social influence mechanisms, even from nonpolarized initial states. Further analytical study of the polarization phenomenon is needed to provide theoretical support to our simulation findings.  Importantly, novel mathematical tools should be developed in order to study stability properties of the polarized equilibria, possibly inspired by the methods proposed in~\cite{zino2020coevolutionary}. Third, effort should be placed to provide a general characterization of the Nash equilibria of the system (besides polarized equilibria) and, if multiple stable equilibria are present, to understand how the network topology and the model parameters affect their basins of attraction. Fourth, several other complex social phenomena might be tackled using our flexible model framework, from diffusion of innovation~\cite{Martins2009innovation,montanari2010spread_innovation,young2011dynamics} to the adoption of social conventions and behaviors~\cite{peytonyoung2015social_norms,Amato2018}. Finally, our model relies on the assumption that individuals make decisions by optimizing a payoff function. Besides being grounded in the social psychology literature and in existing models for decision-making and opinion dynamics, the complexity of the payoff function calls for
a  validation of the proposed model against empirical data, similar to~\cite{ye2021nat}. This, together with the development of techniques to calibrate its parameters will be crucial to enable its practical use in real-world problems.

\appendix

\subsection{Proof of Proposition~\ref{prop:payoff}}\label{app:proof_Payoff_Prop}
 Note that as $\zeta_a\in\{-1,+1\}$, we can determine the maximizers by simply comparing the maximum values obtained for $f_i$ subject to $\zeta_a=-1$ and $\zeta_a=+1$, respectively. To begin, we consider the case $\zeta_a=+1$. Computing the derivative of $f_i(+1,\zeta_o)$ with respect to $\zeta_o$, we obtain
\begin{equation} \begin{aligned}
f_i'(+1,\zeta_o)=&-(1-\lambda_i)\beta_i(1-\gamma_i)\zeta_o\sum\nolimits_{j\in\mathcal{V}}w_{ij}  
\\&+(1-\lambda_i)\beta_i(1-\gamma_i)\sum\nolimits_{j\in\mathcal{V}}w_{ij}y_j \\&-(1-\lambda_i)\beta_i \gamma_i(\zeta_o-u_i)-\lambda_i\beta_i(\zeta_o-1).
\end{aligned}\end{equation} 
The second derivative of $f_i(+1,\zeta_o)$ with respect to $\zeta_o$ is $
f_i''(+1,\zeta_o)=-\beta_i < 0
$, which implies that  $f_i(+1,s)$ is  strictly concave, and hence has a unique maximum.
By using Fermat's (interior extremum) theorem, we solve $f_i'(+1,\zeta_o)=0$ and obtain the following maximizer:
\begin{equation} \label{eq:S}
\zeta^+_o=(1-\lambda_i)(1-\gamma_i)\sum_{j\in\mathcal{V}}w_{ij}y_j+(1-\lambda_i)\gamma_i u_i+\lambda_i.
\end{equation} 
We are left with verifying that the maximizer belongs to the opinion domain, that is, $\zeta^+_o\in[-1,+1]$. Since $\lambda_i\in[0,1]$, $\gamma_i\in[0,1]$, and $\mat W$ is stochastic, we have that $\zeta_o$ in \eqref{eq:S} is a convex combination of $u_i\in[-1,1]$ and $y_j\in[-1,1],\; \forall j\in\mathcal{V}$, and $+1$. It follows that $\zeta^+_o\in[-1,+1]$.

Next, consider $\zeta_a=-1$. Let $\zeta^-_o$ be the maximizer of \eqref{eq:f_function}  subject to $\zeta_a=-1$. Similar to the above, we differentiate $f_i(-1, \zeta_o)$ with respect $\zeta_o$, and by solving $f'_i(-1,\zeta_o)=0$ for $\zeta_o$, we obtain
\begin{equation} \label{eq:barS}
\zeta^-_o=(1-\lambda_i)(1-\gamma_i)\sum_{j\in\mathcal{V}}w_{ij}y_j+(1-\lambda_i)\gamma_i u_i-\lambda_i
\end{equation} 
which can be shown to belong to the domain $\zeta^-_o\in[-1,+1]$, similar to $\zeta^+_o$.

Now, we are left with comparing the two maxima. Define 
\begin{equation} \label{eq:h}
h=(1-\lambda_i)(1-\gamma_i)\sum\nolimits_{j\in\mathcal{V}}w_{ij}y_j+(1-\lambda_i)\gamma_i u_i.
\end{equation} 
Then, we can write $\zeta_o^+=h+\lambda_i$ and $\zeta_o^-=h-\lambda_i$. Using the definitions of $\zeta_o^+$ and $\zeta_o^-$ in \eqref{eq:S} and \eqref{eq:barS}, we  compute
\begin{align}\label{eq:fDif_ssb}
f_i(+1,&\zeta^+_o)-f_i(-1,\zeta_o^-)=\nonumber\\&-2(1-\lambda_i)(1-\gamma_i)\lambda_i\beta_i\sum\nolimits_{j\in \mathcal{V}} w_{ij}(h-y_j)
\nonumber\\&-2(1-\lambda_i)\gamma_i\lambda_i\beta_i(h-u_i)+2\lambda_i h\beta_i(1-\lambda_i)
\nonumber\\&+\frac{(1-\beta_i)\lambda_i}{2}\sum\nolimits_{j\in\mathcal{V}} a_{ij}\big[2x_j+\alpha(1+x_j)\big]
\end{align}

Substituting the expression for $h$ in \eqref{eq:h} into the right hand side of \eqref{eq:fDif_ssb} and simplifying, we obtain:
\begin{equation}
\begin{aligned}
&f_i(1,\zeta^+_o)-f_i(-1,\zeta^-_o)=2\lambda_i\beta_i(1-\lambda_i)(1-\gamma_i)\sum_{j\in \mathcal{V}}w_{ij}y_j\\&+2\lambda_i\beta_i(1-\lambda_i)\gamma_iu_i+\frac{(1-\beta_i)\lambda_i}{2}\sum_{j\in \mathcal{V}} a_{ij}\big[2x_j+\alpha(1+x_j)\big].
\end{aligned}
\end{equation} 
This immediately yields that $\delta_i(\vect{z})=f_i((+1,\zeta^+_o)\vect{z})-f_i((-1,\zeta^-_o),\vect{z})$, where $\delta_i$ was defined in \eqref{eq:delta}. The quantity $\delta_i(\vect{z})$ being positive or negative is therefore equivalent to the pair $(+1,\zeta^+_o)$ or $(-1,\zeta^-_o)$, respectively, being the unique maximizer of $f_i$ for given $\vect z$. The quantity being zero is equivalent to $f_i$ having both as maximizers.
\endproof

\subsection{Proof of Lemma~\ref{lem:potentialGame} and Corollary~\ref{cor:potential_LB}}\label{app:Lemma1}
From Assumption~\ref{as:uniform}, we have $w_{ij}=w_{ji}$ and $a_{ij}=a_{ji}$ for all $i,j\in\mathcal{V}$. Combining this with the fact that all parameters are homogeneous, we can consider a generic $i\in\mathcal{V}$ and write
\begin{align}\label{eq:phi_I}
\Phi&(\vect{z})=\sum\nolimits_{ j\in\mathcal{V}\setminus\{i\}} \eta a_{ij}\Big[(1+\alpha)(1+x_j)(1+x_i)\nonumber\\&\;\;+(1-x_i)(1-x_j)\Big]+\sum\nolimits_{k\in\mathcal{V}\backslash\{i\}}\sum\nolimits_{l\in\mathcal{V}\backslash\{i\}} \eta a_{kl}\cdot
\nonumber\\&\qquad\cdot\Big[(1+\alpha)(1+x_l)(1+x_k)+(1-x_k)(1-x_l)\Big]
\nonumber\\&-\frac{1}{2}\sum_{j\in\mathcal{V}}w_{ij}(y_i-y_j)^2
-\frac{1}{4}\sum_{k\in\mathcal{V}\backslash\{i\}}\sum_{l\in\mathcal{V}\backslash\{i\}}w_{kl}(y_k-y_l)^2
\nonumber\\&-\frac{1}{2}\sum\nolimits_{k\in\mathcal{V}}\frac{\gamma}{(1-\gamma)}(y_k-u_k)^2
\nonumber\\&-\frac12\sum\nolimits_{k\in\mathcal{V}}\frac{\lambda}{(1-\lambda)(1-\gamma)}(y_k-x_k)^2.
\end{align}

Next, we consider two states $\vect{z}$ and $\vect{z}'$ that differ only in the $i$th component, with $\vect{z_i}=\vect{\zeta}=(\zeta_a,\zeta_o)$ and $\vect{z_i}'=\vect{\zeta}'=(\zeta_a',\zeta_o')$. Then, using \eqref{eq:phi_I}, we compute
\begin{align}\label{eq:potential_difference}
\Phi&(\vect{z})-\Phi(\vect{z}')=\sum_{ j\in\mathcal{V}\setminus\{i\}} \eta a_{ij}\big[2x_j+\alpha(1+x_j)\big](\zeta_a-\zeta_a')\nonumber
\\&-\frac{1}{2}\sum\nolimits_{j\in\mathcal{V}\setminus\{i\}}w_{ij}\big[(\zeta_o-y_j)^2-(\zeta_o'-y_j)^2\big]\nonumber
\\&-\frac{1}{2}\frac{\gamma}{(1-\gamma)}\big[(\zeta_o-u_i)^2-(\zeta_o'-u_i)^2\big]\nonumber
\\&-\frac12\frac{\lambda}{(1-\lambda)(1-\gamma)}\big[(\zeta_a-\zeta_o)^2-(\zeta_a'-\zeta_o')^2\big].
\end{align}

Next, we evaluate $f_i(\vect{\zeta},\vect{z}) - f_i( \vect{\zeta}',\vect{z})$, obtaining
\begin{align}\label{eq:02}
    f_i&(\vect{\zeta},\vect{z})-f_i(\vect{\zeta}',\vect{z})= \nonumber\\
      &\frac{\lambda (1-\beta)}{4}\sum\nolimits_{j\in\mathcal{V}\setminus\{i\}}  a_{ij}[2x_j+\alpha(1+x_j)](\zeta_a-\zeta_a')\nonumber \\
    &+\frac{\lambda (1-\beta)}{4} a_{ii}[2x_i+\alpha(1+x_i)](\zeta_a-\zeta_a')\nonumber \\
    &-\frac{1}{2}\beta(1-\lambda)(1-\gamma)\sum_{j\in\mathcal{V}\setminus\{i\}}w_{ij}[(\zeta_o-y_j)^2-(\zeta_o'-y_j)^2]\nonumber \\
    &-\frac{1}{2}\beta(1-\lambda)\gamma[(\zeta_o-u_i)^2-(\zeta_o'-u_i)^2] \nonumber \\
    &-\frac12\lambda\beta[(\zeta_a-\zeta_o)^2-(\zeta_a'-\zeta_o')^2].
\end{align}

 Let $[\beta(1-\lambda)(1-\gamma)]^{-1} = \xi > 0$. For the case $\zeta_a=\zeta_a'$, we have 
$
    \Phi(\vect{z})-\Phi(\vect{z}')=   \xi(f_i(\vect{\zeta},\vect{z})-f_i(\vect{\zeta}',\vect{z}))
$, from which the statement of the lemma follows. Next, consider the case where $\zeta_a\neq\zeta_a'$, and without loss of generality, let $\zeta_a'= -1$ and $\zeta_a = +1$ and $x_i = -1$. Then, the last term on the right hand side of \eqref{eq:02} is given by $-\lambda(1-\beta)a_{ii}$. It follows from \eqref{eq:potential_difference} and \eqref{eq:02} that $\Phi(\vect{z})-\Phi( \vect{z}') -\xi\lambda(1-\beta)a_{ii} = \xi(f_i(\vect{\zeta},\vect{z})-f_i(\vect{\zeta}',\vect{z}))$. 
Observe then, that $f_i(\vect{\zeta},\vect{z})-f_i(\vect{\zeta}',\vect{z})> 0$ implies \begin{equation}\label{eq:bound_increase1}
 \Phi(\vect{z})-\Phi( \vect{z}')  >\xi\lambda(1-\beta)a_{ii} > 0.
\end{equation}

Suppose now, that $\zeta_a' = +1$, $\zeta_a = -1$ and $x_i = +1$. Then, the last term on the right of \eqref{eq:02} is given by $-\lambda(1-\beta)a_{ii}(1+\alpha)$. It follows from \eqref{eq:potential_difference} and \eqref{eq:02} that
\begin{equation}
   \Phi(\vect{z})-\Phi( \vect{z}') -\xi\lambda(1-\beta)a_{ii}(1+\alpha)= \xi(f_i(\vect{\zeta},\vect{z})-f_i(\vect{\zeta}',\vect{z})).
\end{equation}
Observe then, that $f_i(\vect{\zeta},\vect{z})-f_i(\vect{\zeta}',\vect{z}) > 0$ implies 
\begin{equation}\label{eq:bound_increase2}
  \Phi(\vect{z})-\Phi( \vect{z}')   >\xi\lambda(1-\beta)a_{ii}(1+\alpha)  > 0\,,
\end{equation}
which completes the proof of Lemma~\ref{lem:potentialGame}. Finally, the minimum of \eqref{eq:bound_increase1} and \eqref{eq:bound_increase2} over all $i\in\mathcal{V}$ gives the bound in Corollary~\ref{cor:potential_LB}, by noting $\alpha \geq 0$.
\endproof

\subsection{Proof of Lemma~\ref{lem:opinion_convergence}}\label{app:Lemma2}

We first prove there is a unique $\vect y^*$ that is the equilibrium of the opinion dynamics in \eqref{eq:yUpdate} for a given (fixed) $\vect x^*$. Without loss of generality, assume that $\vect x^* = [\vect 1_r^\top, \vect 1_{n-r}^\top]^\top$ for some $0\leq r \leq n$. Observe from \eqref{eq:yUpdate} that any equilibrium $\vect y^*$ must have entries $y_i^*$ such that
\begin{equation}
    y_i^* = (1-\lambda)\big[(1-\gamma)\sum\nolimits_{j\in\mathcal{V}} w_{ij} y_j^* + \gamma u_i\big] + \lambda x_i^*
\end{equation}
or in vector form $\vect y^* = (1-\lambda)[(1-\gamma)\mat{W}\vect y^*+\gamma\vect u] +\lambda\vect x^*$, 
where $\vect u = [u_1, \hdots, u_n]^\top$. Rearranging for $\vect y^*$ yields
$[\mat I - (1-\lambda)(1-\gamma)\mat W]\vect y^* = (1-\lambda)\gamma\mat{u} + \lambda\mat{x}^*$. 
Since $\mat W$ is stochastic and irreducible, the Perron--Frobenius Theorem~\cite{berman1994nonnegative} establishes that $\mat W$ has a simple eigenvalue equal to 1, with all other eigenvalues having modulus no greater than 1. It follows that the spectral radius of $(1-\lambda)(1-\gamma)\mat W$ is strictly less than 1, and thus $\mat I - (1-\lambda)(1-\gamma)\mat W$ is invertible. Thus,
$\vect y^* = [\mat I - (1-\lambda)(1-\gamma)\mat W]^{-1}[(1-\lambda)\gamma\mat{u} + \lambda\mat{x}^*]$ is unique.

Next, we prove convergence to $\vect y^*$, given the fixed action vector $\vect x^*$. Using a compact vector form, we verify that the potential function $\Phi(\vect z)$ can be expressed in quadratic form as
\begin{align}
    \Phi(\vect{z})& =-\frac12\big(\vect{y}^\top (\mat{I}-\mat{W}+\mat{L}+\mat{G}) \vect{y}-2\vect{y}^\top\mat{G} \vect{u}-2\vect{y}^\top\mat{L} \vect{x}^*\nonumber \\
    &\qquad +(\vect{x}^*)^\top\mat{L} \vect{x}^*+\vect{u}^\top\mat{G} \vect{u}\big)+\mathcal{F}(\vect{x}^*)
\end{align}
where $\mat{L}=\frac{\lambda}{(1-\lambda)(1-\gamma)}\mat{I}$, $\mat{G}=\frac{\gamma}{(1-\gamma)} \mat I$, and 
\begin{equation*}
\small\mathcal{F}(\vect{x}^*)=\sum_{i\in\mathcal{V}}\sum_{j\neq i} \eta \frac{a_{ij}}{2}\big[(1+\alpha)(1+x_j^*)(1+x_i^*)+(1-x_i^*)(1-x_j^*)\big].
\end{equation*}
According to~\cite{berman1994nonnegative}, $(\mat{I}-\mat{W}+\mat{L}+\mat{G})$ is positive definite for any $\lambda\neq 0$ and $\gamma\neq 0$ (c.f. Chapter~6, and the definition of $M$-matrices). Therefore, $\Phi(\vect z)=\Phi([\vect x,\vect y])$ is  continuous in $\vect y$,  and, for fixed $\vect x=\vect x^*$,  bounded from above and from below, i.e., $ \underline{\Phi}\leq \Phi([\vect x^*,\vect y])\leq\bar \Phi$, where $\bar \Phi$ and $\underline{\Phi}$ represent the value of $\Phi([\vect x,\vect y])$ at the global maximum and global minimum with respect to $\vect y$, constrained to $\vect x=\vect x^*$, respectively. 

In Lemma~\ref{lem:potentialGame}, we proved that the coevolutionary game is a generalized ordinal potential game with potential function $\Phi$. For convenience, denote $\Phi(t):=\Phi(\vect z(t))=\Phi([\vect x^*,\vect y(t)])$. By using $\vect z = \vect z(t+1)$ and $\vect z^\prime = \vect z(t)$ in \eqref{eq:gopg_ineq}, it follows that under the best-response update in \eqref{eq:br_xy}, there holds $\Phi(t+1) - \Phi(t) > 0$ if agent $i$ is active at time $t$ and $y_i(t+1) \neq y_i(t)$. It follows from Assumption~\ref{as:activation} that $\Phi(t+T) - \Phi(t) \geq 0$ for all $t \geq 0$ with equality if and only if $\vect y(t+T) = \vect y(t)$. In other words, $\Phi(t)$ must strictly increase over any time window of period $T$ unless $y_i(t+T) = y_i(t)$ for all $i\in\mathcal{V}$. This is because every agent activates at least once inside the time window, and thus either at least one agent $i$ changes its opinion $y_i(t+T) \neq y_i(t)$ or every agent has an unchanged opinion over that time window. In the latter case, clearly the opinions have reached an equilibrium, and the results above establish there is a unique equilibrium vector $\vect y^*$.

This implies that the sequence $\Phi_t \triangleq \{\Phi(t)\}_{t=0}^\infty$ is monotonically increasing across every time period $T$, and as we established earlier, it is bounded from above by the unique global maximum $\bar \Phi$. It follows from  the monotone convergence theorem for sequences, that $\Phi_t$ converges asymptotically to $\bar \Phi$, and hence $\lim_{t\to\infty} \vect y(t) = \vect y^*$ asymptotically.
\endproof

\subsection{Proof of Theorem~\ref{th:convergence}}\label{app:Theorem1}

 The proof centers around the potential function defined in \eqref{eq:phi}, and its properties established in Lemma~\ref{lem:potentialGame} and Corollary~\ref{cor:potential_LB}. We establish convergence by focusing on proving that actions must converge to a steady state $\vect x^*$ in a finite number of time steps. Letting $\tau$ denote the time step for which the action has converged, that is, $\vect x(t+1) = \vect x(t) = \vect x^*$ for all $t\geq \tau$, convergence of the opinion vector $\vect y(t)$ follows immediately from Lemma~\ref{lem:opinion_convergence}.

Note that $\Phi(\vect z)$ is bounded from above and from below, and let us denote the maximum and minimum value of $\Phi(\vect z)$ by $\bar\Phi$ and $\underline\Phi$, respectively. With an abuse of notation, we write the potential function as an explicit function of the time, that is $\Phi(t)=\Phi(\vect{z}(t))$. Assume without loss of generality that node $i\in\mathcal{V}$ is active at time $t$. Then, $\vect{z}(t+1)$ may differ from $\vect{z}(t)$ only in the $i$th row. Now, if individual $i$ chooses to change their action, i.e. $x_i(t+1)\neq x_i(t)$, then we know from \eqref{eq:PhiDifPositive} in Corollary~\ref{cor:potential_LB}  that the difference in potential is bounded from below by a positive constant, i.e., $\Phi(t+1)-\Phi(t)>c>0$. 

Recall from Assumption~\ref{as:activation}, that every individual will activate at least once in every time window
$[t; t + T), \; t \in \mathbb Z_+$. Thus, there can be at most $\lfloor \frac{\bar \Phi - \underline{\Phi}}{c}\rfloor$ time instants where the active agent $i$ has $x_i(t+1) \neq x_i(t)$ for $t < \bar T$ for some $\bar T \in \{\mathbb{Z}_+ \cup \infty\}$. If $\bar T$ is finite, then convergence to a maximum of $\Phi$ (though not necessarily the global maximum) must occur and consequently $\vect x(t)$ converges to some $\vect x^*$ associated with the maximum in finite time, before $\bar T$.

If $\bar T = \infty$ on the other hand, this would imply that there exist an infinitely long sequence of agent activations in which $x_i(t+1) = x_i(t)$ for every active agent $i$, during which the opinions converge asymptotically to some steady state vector $\vect{y}^*$ as detailed in Lemma~\ref{lem:opinion_convergence}, and then there exists an agent $j$ such that $\lim_{t\to\infty} x_j(t+1) \neq x_j(t)$. That is, an agent chooses to change their action after a long time period in which no action has changed, while the opinions have reached to a neighborhood of the opinion equilibrium. We are now going to show such an infinitely long sequence cannot exist, by showing that if there is a sufficiently long but finite sequence with no action change, then it is not possible for any future action change for any agent after this sequence.

To establish the contradiction, suppose that there is a finite sequence of agent activations with no change in the actions, except agent~$i$ who activates at the end of this sequence and changes action. Let $t_2$ be the time step at end of this sequence, and let $t_1$, with $t_1 < t_2$, be the supremum time step inside this sequence at which agent~$i$ was last active before $t_2$. Without loss of generality, assume $x_i(t_1) = x_i(t_1 + 1) = -1$ which implies 
$x_i(t_2+1) = +1 \neq x_i(t_2) = x_i(t_1+1) = -1$.
Now, \eqref{eq:S_delta} implies that $\delta_i( \vect{z}(t_1)) \leq 0$ while $\delta_i ( \vect{z}(t_2)) > 0$. We are going to show that $\delta_i ( \vect{z}(t_2))$ being positive creates a contradiction.

First, we consider the case  $\delta_i(\vect{z}(t_1))=0$, which implies
\begin{align}\label{eq:deltaDif0}
   &2\lambda\beta\Big[(1-\lambda)(1-\gamma)\sum\nolimits_{j\in \mathcal{V}}w_{ij}y_j(t_1)+(1-\lambda)\gamma u_i\Big]\nonumber\\&+\frac{(1-\beta)\lambda}{2}\sum_{j\in\mathcal{V}} a_{ij}\big[2x_j(t_1)+\alpha(1+x_j(t_1))\big]=0.
\end{align}
Rearranging \eqref{eq:yUpdate} and substituting it into the first term of \eqref{eq:deltaDif0} with $x_i(t_1+1)=-1$, we can write \eqref{eq:deltaDif0} as
    \begin{align}\label{eq:deltaDif0T1}
   &2\lambda\beta\big[y_i(t_1+1)+\lambda\big]\nonumber\\&+\frac{(1-\beta)\lambda}{2}\sum_{j\in\mathcal{V}} a_{ij}\big[2x_j(t_1)+\alpha(1+x_j(t_1))\big]=0.
    \end{align}
     Now, positivity of $\delta_i ( \vect{z}(t_2))$ implies
    \begin{align}\label{eq:deltaDif00}
    &2\lambda\beta(1-\lambda)(1-\gamma)\sum\nolimits_{j\in \mathcal{V}}w_{ij}y_j(t_2)+2\lambda\beta(1-\lambda)\gamma u_i\nonumber\\&+\frac{(1-\beta)\lambda}{2}\sum_{j\in\mathcal{V}} a_{ij}\big[2x_j(t_2)+\alpha(1+x_j(t_2))\big]>0.
    \end{align}
    Similarly, by recalling \eqref{eq:yUpdate} and with $x_i(t_2+1)=+1$, we can write \eqref{eq:deltaDif00} as follows:
        \begin{align}\label{eq:deltaDif0T2}
    &2\lambda\beta\big[y_i(t_2+1)-\lambda\big]\nonumber\\&+\frac{(1-\beta)\lambda}{2}\sum_{j\in\mathcal{V}} a_{ij}\big[2x_j(t_2)+\alpha(1+x_j(t_2))\big]>0.
    \end{align}
By hypothesis, $x_j(t_2) = x_j(t_1)$ for all $j\neq i$, which implies that $\vect x(t_2) = \vect x(t_1)$. Thus, we can rearrange \eqref{eq:deltaDif0T1} and substitute  it into the left hand side of \eqref{eq:deltaDif0T2} to obtain $2\lambda\beta[y_i(t_2+1)-\lambda-y_i(t_1+1)-\lambda]>0$. 
By taking this sequence to be sufficiently long, then Lemma~\ref{lem:opinion_convergence} guarantees asymptotic convergence of $\vect y(t)$ to $\vect y^*$, and thus $\vect y(t_2+1)$ and $\vect y(t_1+1)$ are both in a small neighborhood of $\vect y^*$. Further, we have $y_i(t_1+1)=y_i^*+ \hat \varepsilon_i^{t_1}$ and $y_i(t_2+1)=y_i^*+\hat\varepsilon_i^{t_2}$ 
with $| \hat\varepsilon_i^{t_2}|\leq |\hat\varepsilon_i^{t_1}|$. This allows us to write
        \begin{equation} \label{eq:deltaDif0F}\begin{aligned}
    &2\lambda\beta\big[\hat\varepsilon_i^{t_2}-\hat\varepsilon_i^{t_1}-2\lambda\big]>0.
        \end{aligned}
    \end{equation} 
Noting that $\hat\varepsilon_i^{t_2}-\hat\varepsilon_i^{t_1}\leq 2|\hat\varepsilon_i^{t_1}|$, it follows that for a sufficiently small $\hat\varepsilon_i^{t_1}$ (as occurs when there is a sufficiently long sequence of agent activations with no actions changing), the left of \eqref{eq:deltaDif0F} cannot be positive, which results in a contradiction.

Now, we consider the case   $\delta_i(\vect{z}(t_1)) < 0$. Again, by hypothesis, $x_j(t_2) = x_j(t_1)$ for all $j\neq i$. Note that this implies that $\vect x(t_2) = \vect x(t_1)$.  Using this and adding and subtracting $ 2\lambda\beta(1-\lambda)(1-\gamma)\sum_{j\in \mathcal{V}}w_{ij}y_j(t_1)$  to the left hand side of \eqref{eq:deltaDif00}, we obtain
\begin{equation} \label{eq:deltaDif}
2\lambda\beta(1-\lambda)(1-\gamma)\sum_{j\in \mathcal{V}}w_{ij}(y_j(t_2)-y_j(t_1))+\delta_i(\vect{z}(t_1))>0
\end{equation} 
Similar to the above, if the sequence in question is sufficiently long, then Lemma~\ref{lem:opinion_convergence} guarantees asymptotic convergence of $\vect y(t)$ to $\vect y^*$, and also that $\vect y(t_1)$ and $\vect y(t_2)$ are in a small neighborhood of $\vect y^*$. Moreover, $y_i(t_1)=y_i^*+ \varepsilon_i^{t_1}$
and
$y_i(t_2)=y_i^*+\varepsilon_i^{t_2}$
with $| \varepsilon_i^{t_2}|\leq |\varepsilon_i^{t_1}|$.
Hence, we can write \eqref{eq:deltaDif} as
\begin{equation} \label{eq:DeltaDif2}
2\lambda\beta(1-\lambda)(1-\gamma)\sum_{j\in \mathcal{V}}w_{ij}(\varepsilon_j^{t_2}-\varepsilon_j^{t_1})+\delta_i(\vect{z}(t_1))>0
\end{equation} 
Note that $\delta_i(\vect{z}(t_1))<0$ and 
$
\sum_{j\in \mathcal{V}}w_{ij}(\varepsilon_j^{t_2}-\varepsilon_j^{t_1})\leq 2|\sum_{j\in \mathcal{V}}w_{ij}\varepsilon_j^{t_1}|
$. As a result, for a sufficiently small $\varepsilon_j^{t_1}$, which is guaranteed if the sequence is sufficiently long, the left hand side of \eqref{eq:DeltaDif2} can never be positive. Therefore, $\delta_i(\vect{z}(t_2))>0$ cannot hold, which implies that the action $x_i(t_2+1)$ in fact does not change from $x_i(t_2)$. 

To conclude, we showed that there cannot exist an infinitely long sequence of agent activation such that the actions do not change, while the opinions are converging asymptotically to its unique equilibrium $\vect y^*$. This implies that the actions change at most $\lfloor \frac{\bar \Phi - \underline{\Phi}}{c}\rfloor$ times in a finite number of time steps; with the actions converging in finite time, asymptotic convergence of the opinions follows from Lemma~\ref{lem:opinion_convergence}. 
\endproof

\subsection{Proof of Theorem~\ref{th:polarized}}\label{app:polarized}

 The proof is comprised of two parts. First, we prove that given a fixed action vector $\vect{x}^*$ that is polarized with respect to a partitioning $(\mathcal{V}_p,\mathcal{V}_n)$, as defined in Definition~\ref{def:polarized_state}, and under the conditions in \eqref{eq:pol_y_cond}, the unique vector $\vect{y}^*$ that satisfies the equilibrium conditions for \eqref{eq:yxUpdate_POL} is polarized. Second, we show that conditions in \eqref{eq:pol_xp_cond} and \eqref{eq:pol_xn_cond} guarantee that the action vector $\vect{x}^*$ in Part 1 is invariant under \eqref{eq:yxUpdate_POL}, which implies that $\vect{z}^*$ is a polarized equilibrium of the coevolutionary model.

\textit{Part 1:} Let $\vect{x}^*$ be an action vector that is polarized with respect to a partitioning $(\mathcal{V}_p,\mathcal{V}_n)$. Define $y_m^+=\min_{i\in\mathcal{V}_p} y_i^*$ and $y_m^-=\min_{i\in\mathcal{V}_n} y_i^*$, where $\vect{y}^*$ is the unique equilibrium opinion vector if $\vect{x}^*$ was fixed (see Lemma~\ref{lem:opinion_convergence}). From \eqref{eq:yxUpdate_POL}, the equilibrium evaluates to be
\begin{equation}
    y^*_i = \lambda\mathcal{S}(\delta_i(\vect{z})) + (1-\lambda)\bigg(\sum_{j\in\mathcal{V}_p} w_{ij} y_j^* + \sum_{k\in\mathcal{V}_n} w_{ik} y_k^*\bigg).
\end{equation}
It follows that $y_m^+\geq y_m^-$ and
\begin{equation} \label{eq:yPlusInEq}
y_m^+\geq \lambda+(1-\lambda)\big[\underline d_py_m^++(1-\underline d_p)y_m^-\big]
\end{equation} 
and 
\begin{equation} \label{eq:yMinusInEq}
y_m^-\geq -\lambda+(1-\lambda)\big[\bar d_ny_m^-+(1-\bar d_n)y_m^+\big].
\end{equation} 
Rearranging \eqref{eq:yMinusInEq} and substituting in \eqref{eq:yPlusInEq}, and then solving for $y_m^+$ yields
\begin{equation} \label{eq:yPlusInEqF}
y_m^+\geq \lambda \frac{1-(1-\lambda)(1+\bar d_n-\underline d_p)}{1-(1-\lambda)(\bar d_n+\underline d_p)-(1-\lambda)^2(1-\bar d_n-\underline d_p)}.
\end{equation} 
Expanding and simplifying the denominator on the right-hand-side of \eqref{eq:yPlusInEqF} yields $
\lambda\left(1+(1-\lambda)(1-\bar d_n - \underline d_p)\right)$. 
As $\bar d_n < 1$ by definition, it follows that $    1+(1-\lambda)(1-\bar d_n - \underline d_p) > 1-(1-\lambda)\underline d_p$. 
Further, $\underline d_p < 1$ by definition, which implies $(1-\lambda)\underline d_p < 1$. Thus, $1+(1-\lambda)(1-\bar d_n - \underline d_p) > 1-(1-\lambda)\underline d_p > 0$. As the denominator on the right-hand-side of \eqref{eq:yPlusInEqF} is positive, the right-hand-side of \eqref{eq:yPlusInEqF}  is positive  if and only if $1-(1-\lambda)(1+\bar d_n-\underline d_p)>0$ or
\begin{equation} \label{eq:yPlus_cond}
\lambda> {(\bar d_p-\underline d_n)}/{(1+\bar d_p-\underline d_n)}.
\end{equation} 
That is, $y_i^* > 0$ for all $i\in \mathcal{V}_p$ if \eqref{eq:yPlus_cond} holds.
Using analogous arguments, but for the quantities $y_M^+ = \max_{i\in\mathcal{V}_p} y_i^*$ and $y_M^- = \max_{i\in\mathcal{V}_n} y_i^*$, it follows that $y_i^* < 0,\forall\,i \in \mathcal{V}_n$ if
\begin{equation} \label{eq:yMin_cond}
\lambda>{(\bar d_n-\underline d_p)}/{(1+\bar d_n-\underline d_p)}.
\end{equation} 
This secures \eqref{eq:pol_y_cond}.

\textit{Part 2:} We now show that Eqs.~(\ref{eq:pol_xp_cond})--(\ref{eq:pol_xn_cond}) are sufficient for the $\vect{x}^*$ polarized with respect to the partitioning $(\mathcal{V}_p, \mathcal{V}_n)$ to be an equilibrium.

For all $i\in\mathcal{V}_p$ and $t\geq 0$, $x_i(t) = +1 = x_i(t+1) = +1$ if and only if $\delta_i(\vect{z}(t)) > 0$, which from \eqref{eq:deltaPol} is implied by
\begin{equation} \label{eq:deltaTheoremPol}
2\lambda\beta(1-\lambda)(\underline d_py_m^++(1-\underline d_p)y_m^-)+(1-\beta)\lambda(2\underline d_p-1)>0,
\end{equation} 
where $y_m^-$ and $y_m^+$ were defined in \textit{Part 1} of this proof. Noting \eqref{eq:yxUpdate_POL} and \eqref{eq:yPlusInEq}, and simplifying the term $\lambda$, we write \eqref{eq:deltaTheoremPol} as 
\begin{equation}
2\beta(y_m^+-\lambda)+(1-\beta)(2\underline d_p-1)>0.
\end{equation} 
Substituting $y_m^+$ from \eqref{eq:yPlusInEqF} and simplifying yields \eqref{eq:pol_xp_cond}.

Following a similar argument, for any $i\in\mathcal{V}_n$ and all $t\geq 0$, $x_i(t) = -1$ and $x_i(t+1) = -1$ if and only if $\delta_i(x(t),y(t)) < 0$, which from \eqref{eq:deltaPol} is implied by
\begin{equation} \label{eq:deltaTheoremPol_neg}
2\lambda\beta(1-\lambda)(\underline d_ny_M^-+(1-\underline d_p)y_M^+)-(1-\beta)\lambda(2\underline d_n-1)<0
\end{equation} 
where $y_M^-$ and $y_M^+$ were defined in \textit{Part 1} of this proof. Utilizing again \eqref{eq:yxUpdate_POL}, we can write \eqref{eq:deltaTheoremPol_neg} as
$
2\beta(y_M^-+\lambda)-(1-\beta)(2\underline d_n-1)<0.
$
Finally, substituting the expression for $y^-_M$, computed to derive \eqref{eq:yMin_cond}, we obtain \eqref{eq:pol_xn_cond}.\endproof

\subsection{Proof of Theorem~\ref{th:polarized_convergence}}\label{app:polarized_convergence}

Assuming that $\vect{x}(t)$  and $\vect{y}(t)$ are polarized for a given $t\in\mathbb Z_+$, we start by showing that if 
\begin{equation} \label{eq:xpol_cond}\begin{cases}
\sum_{j\in\mathcal{V}_p}w_{ij}> \frac12\big(1+\frac{\beta(1-\lambda)}{1-\beta\lambda}\big),  &i\in\mathcal{V}_p,
\\
\sum_{j\in\mathcal{V}_n}w_{ij}> \frac12\big(1+\frac{\beta(1-\lambda)}{1-\beta\lambda}\big),  &i\in\mathcal{V}_n
\end{cases}\end{equation} 
holds, then the action vector $\vect{x}(t+1)$ will be polarized; then we show that  if also 
\begin{equation} \label{eq:ypol_cond}\begin{cases}
\sum_{j\in\mathcal{V}_p}w_{ij}> \frac{1-2\lambda}{1-\lambda},  &i\in\mathcal{V}_p,
\\
\sum_{j\in\mathcal{V}_n}w_{ij}> \frac{1-2\lambda}{1-\lambda},  &i\in\mathcal{V}_n
\end{cases}\end{equation} 
holds, then  the opinion vector $\vect{y}(t+1)$ will also be polarized. Hence, by induction, if $\vect{z}(0)$ is polarized, then $\vect{z}(t)$ will be polarized, for all $t\in\mathbb Z_+$. Finally, Theorems~\ref{th:convergence} and~\ref{th:polarized} are leveraged to guarantee convergence to an equilibrium and that such equilibrium is polarized, respectively.

Let us consider an arbitrary time step $t$, at which an agent $i$ activates, and suppose that the system is in a polarized state $\vect{z}(t)=\vect{z}$ with respect to a partition $(\mathcal{V}_p, \mathcal{V}_n)$. We start by showing that if \eqref{eq:xpol_cond} holds, then $x_i(t+1) = x_i(t)$, and thus the action vector $\vect x$ remains polarized. We demonstrate the computations for agent $i\in \mathcal{V}_p$, with the arguments for agent $i\in \mathcal{V}_n$ being almost identical and thus omitted. 

Suppose that the agent active at time $t$ is agent $i\in \mathcal{V}_p$, which implies that $x_i(t) = +1$. Now, agent $i$ will not change action if  $\text{sgn}(\delta_i(\vect z(t)))\geq 0$. Considering the partitioning $(\mathcal{V}_p, \mathcal{V}_n)$, and dropping the time argument $t$ for convenience, we can write \eqref{eq:deltaPol} as:
\begin{equation} \label{eq:deltaPol_01}
   \delta_i(\vect{z})=p\sum_{j\in \mathcal{V}_p}w_{ij}y_j
   +p\sum_{k\in \mathcal{V}_n}w_{ik}y_k+q\Big(\sum_{j\in\mathcal{V}_p} w_{ij}-\frac12\Big)
    \end{equation} 
where $p=2\lambda\beta(1-\lambda)$ and $q=2\lambda(1-\beta)$.
 Since $y_k\in[-1, 0),\; \forall k\in\mathcal{V}_n$ and $y_j\in(0, +1],\; \forall j\in\mathcal{V}_p$, \eqref{eq:deltaPol_01} is implied by
\begin{equation}\label{eq:deltaPol_02}
  \delta_i(\vect{z})
   \geq -p\sum\nolimits_{k\in \mathcal{V}_n}w_{ik}+q\Big(\sum\nolimits_{j\in\mathcal{V}_p} w_{ij}-\frac12\Big).
\end{equation}
We know that $\sum_{k\in \mathcal{V}_n}w_{ik} = 1-\sum_{j\in \mathcal{V}_p}w_{ij}$ because $\mat W$ is stochastic. Thus, \eqref{eq:deltaPol_02} is further implied by 
\begin{equation}
    \delta_i(\vect{z})
   \geq (p+q) \sum\nolimits_{j\in \mathcal{V}_p}w_{ij}-\Big(p+\frac12 q\Big)
\end{equation}
It follows that
\begin{equation}
 \sum\nolimits_{j\in \mathcal{V}_p}w_{ij}\geq \frac12\Big(1+\frac{p}{p+q}\Big)= \frac12\Big(1+\frac{\beta(1-\lambda)}{1-\beta\lambda}\Big)
\end{equation} 
implies $\delta_i(\vect z) \geq 0$, which yields $x_i(t+1) = x_i(t) = +1$. Following a similar path for $i\in\mathcal{V}_n$, we conclude that the action vector at $\vect x(t+1)$ remains polarized with respect to the partitioning $(\mathcal{V}_p, \mathcal{V}_n)$.

Next, we prove that if Eqs.~(\ref{eq:xpol_cond}) and~(\ref{eq:ypol_cond}) holds, then $\text{sgn} (y_i(t+1)) = \text{sgn} (y_i(t))$, and thus $\vect y(t+1)$ is polarized with respect to the same partitioning $(\mathcal{V}_p, \mathcal{V}_n)$ just as $\vect y(t)$.  Again, we consider agent $i\in \mathcal{V}_p$ being the active agent at time $t$, and omit the computations for $i\in \mathcal{V}_n$ as they follow a similar path.

Consider agent $i\in \mathcal{V}_p$ as the active agent at time $t$. Since \eqref{eq:xpol_cond} holds,
then $\delta_i(\vect z(t)) \geq 0$, as detailed above. Then, with $y_k(t)\in[-1,0),\; \forall k\in\mathcal{V}_n$, and $y_j(t)\in(0,+1],\; \forall j\in\mathcal{V}_p$, observe that \eqref{eq:yUpdate} implies that 
\begin{equation} \label{eq:01}\begin{array}{lll}
y_i(t+1)&\geq&(1-\lambda)\big(-\sum_{k\in \mathcal{V}_n}w_{ik} \big)+\lambda\\&=& (1-\lambda)\big( \sum_{j\in \mathcal{V}_p}w_{ij}-1 \big)+\lambda\end{array}
\end{equation}
where the latter holds since $\sum_{j\in \mathcal{V}_n}w_{ij} = 1- \sum_{j\in \mathcal{V}_p}w_{ij}$. Thus, $y_i(t+1) > 0$ is implied by the right-hand side of \eqref{eq:01} being strictly positive, and this holds if the first condition in \eqref{eq:ypol_cond} holds. Note that \eqref{eq:ypol_cond} always holds if $\lambda>\frac12$ because $w_{ij}$ are nonnegative. This, combined with \eqref{eq:xpol_cond}, secures \eqref{eq:condition_polarized}. 

An induction argument can then used to guarantee that $\vect{z}(t)$ will be polarized, for any $t\geq 0$, if $\vect z(0)$ is polarized with respect to the partition $(\mathcal{V}_p, \mathcal{V}_n)$ and \eqref{eq:condition_polarized} holds. Finally,  Theorem~\ref{th:convergence} ---which can be applied since Assumptions~\ref{as:uniform}--\ref{as:activation} hold--- guarantees that convergence whereby $\lim_{t\to\infty} \vect{z}(t)\to\vect{z}^*$. Since $\vect z(t)$ is polarized with respect to $(\mathcal{V}_p, \mathcal{V}_n)$, and because \eqref{eq:condition_polarized} verifies conditions~(\ref{eq:pol_y_cond})--(\ref{eq:pol_xn_cond}) of Theorem~\ref{th:polarized} (which can be applied since Assumptions~\ref{as:uniform}--\ref{as:polarization} hold), then $\vect z^*$ must be polarized with respect to $(\mathcal{V}_p, \mathcal{V}_n)$. 
\endproof

\subsection{Proof of Corollary~\ref{cor:bipartite_consensus}}\label{app:bipartite_consensus}

First, observe that if $d>1/2$, then Eqs.~(\ref{eq:pol_y_cond})--(\ref{eq:pol_xn_cond}) hold, and Theorem~\ref{th:polarized} thus establishes that the system has a polarized equilibrium $\vect z^* = (\vect x^*, \vect y^*)$ with respect to the partition $(\mathcal{V}_p,\mathcal{V}_n)$. According to Lemma~\ref{lem:opinion_convergence}, $\vect y^*$ is unique, and thus we need only to verify that $\vect y^*$ satisfies the expression given in Eqs.~(\ref{eq:uniform1})--(\ref{eq:uniform2}). Under \eqref{eq:uniform}, it is straightforward to verify from \eqref{eq:yxUpdate_POL} (by considering the equilibrium equations) that $y^*$ as defined by \eqref{eq:uniform1} satisfies the system of equations
\begin{equation}
 \left\{   \begin{array}{l}
y^+=(1-\lambda)(d y^++(1-d)y^-)+\lambda\\
y^-=(1-\lambda)(d y^-+(1-d)y^+)-\lambda
\end{array}\right.
\end{equation}
and whose solution is \eqref{eq:uniform2}.\endproof

\begin{IEEEbiography}[{\includegraphics[width=1in,height=1.25in,clip,keepaspectratio]{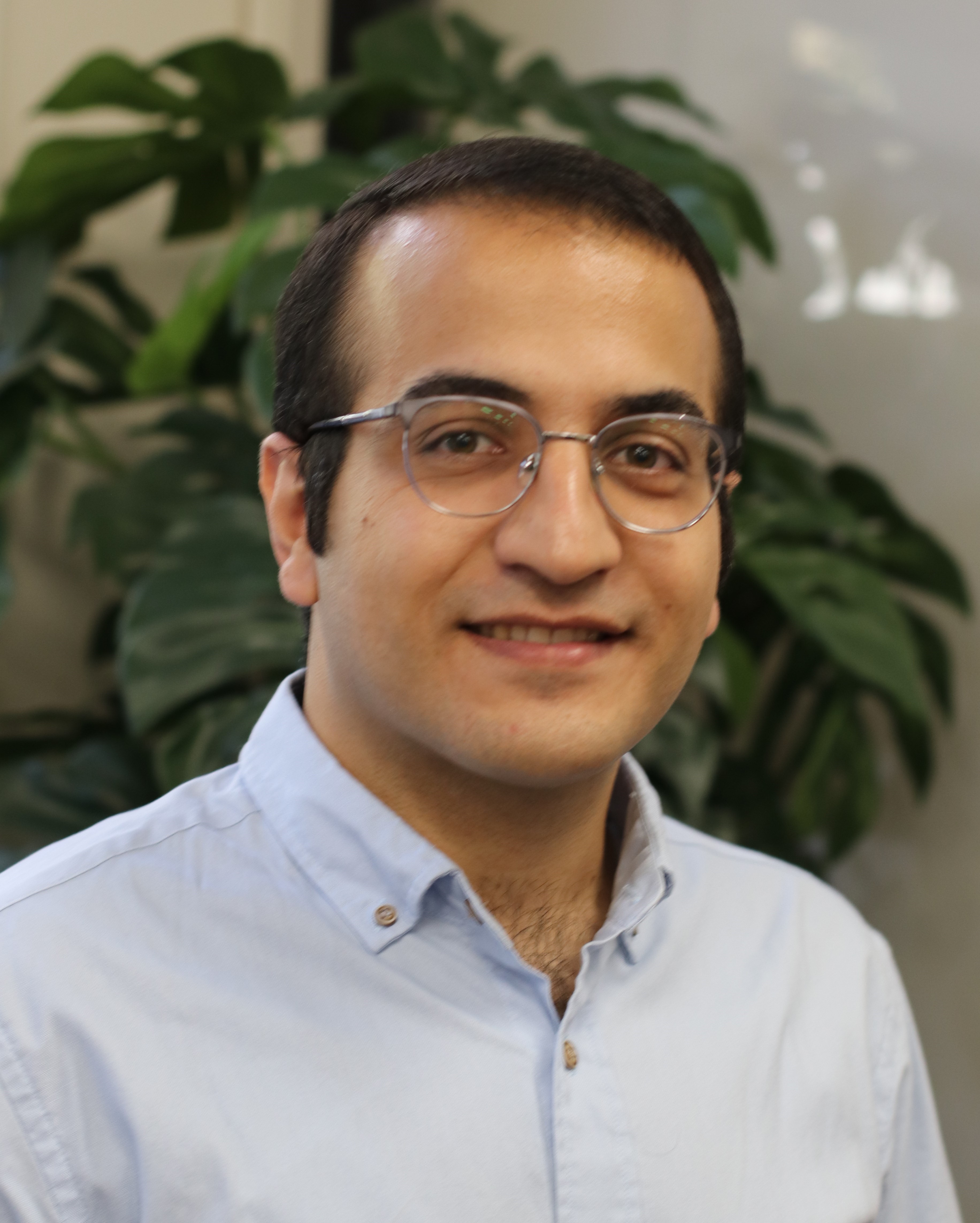}}]{Hassan Dehghani Aghbolagh} was born in Urmia, Iran. He received his B.Sc., and M.Sc. in Electrical Engineering from the Islamic Azad University of Urmia, and the University of Tabriz, Iran, in 2014 and 2016, respectively. He was a member of the Network Control Systems (NCS) Lab at the Department of Electrical Engineering, University of Tabriz. In 2022, he received his Ph.D. degree under the supervision of Professor Zhiyong Chen from the University of Newcastle, Australia.  Currently working as a professional Automation Engineer in the water industry, he is actively seeking research opportunities to contribute to this field.

\end{IEEEbiography}

\begin{IEEEbiography}[{\includegraphics[width=1in,height=1.25in,clip,keepaspectratio]{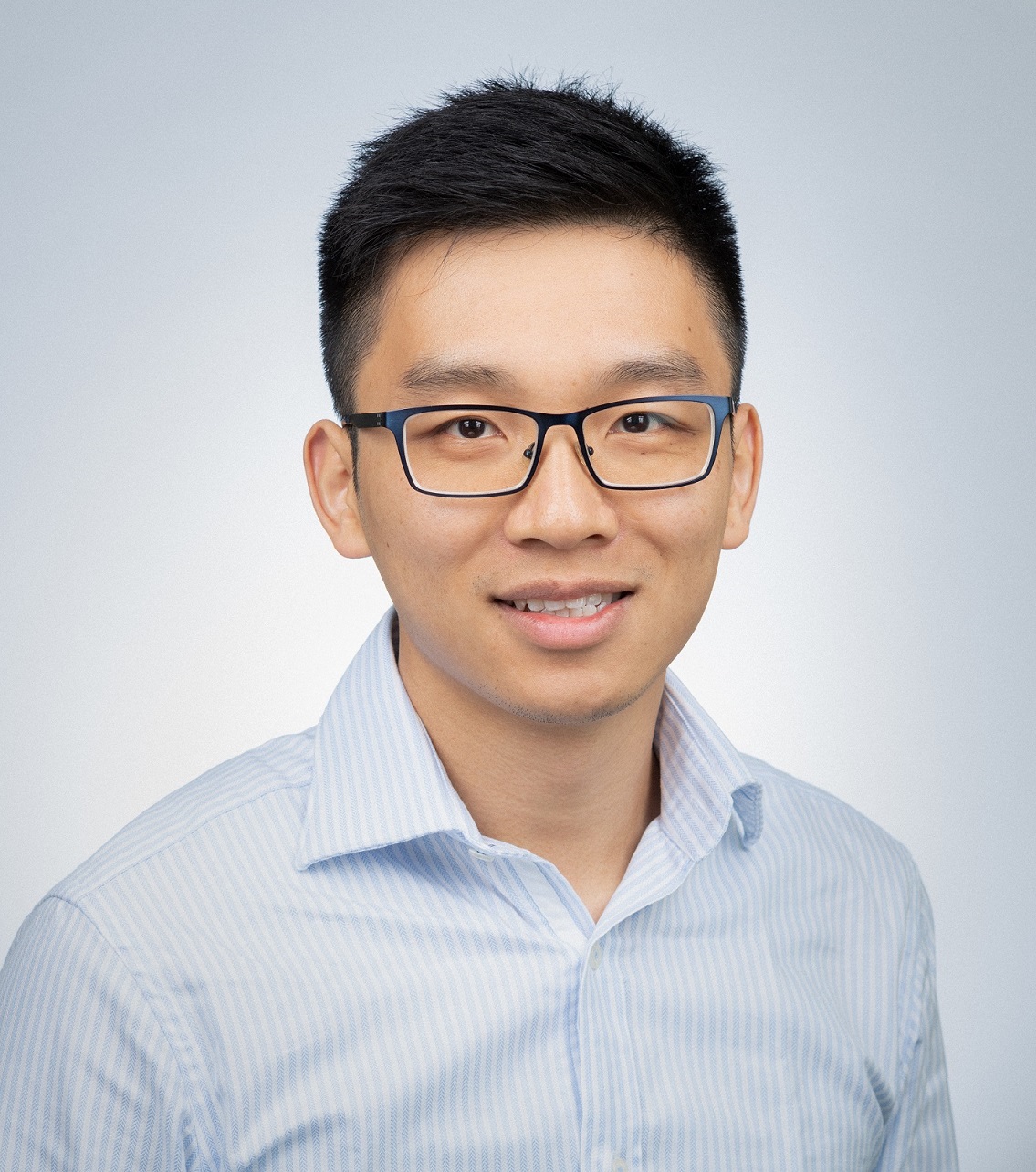}}]{Mengbin Ye} (S’13-M’18) was born in Guangzhou,
China. He received the B.E. degree (with First
Class Honours) in mechanical engineering from the
University of Auckland, Auckland, New Zealand in
2013, and the Ph.D. degree in engineering at the
Australian National University, Canberra, Australia
in 2018. From 2018-2020, he was a postdoctoral
researcher with the Faculty of Science and Engineering, University of Groningen, Netherlands. From 2020 - 2021, he was an Optus Fellow at the Optus–Curtin
Centre of Excellence in Artificial Intelligence, Curtin
University, Perth, Australia. In 2021, he commenced a four year Western Australia Premier's Early to Mid-Career Fellowship, hosted by Curtin University. He was awarded the J.G. Crawford Prize (Interdisciplinary) in 2018, ANU’s premier award recognising graduate research excellence. He has also
received the 2018 Springer PhD Thesis Prize, and was Highly Commended
in the Best Student Paper Award at the 2016 Australian Control Conference.
His current research interests include opinion formation and decision making
in complex social networks, epidemic modelling and control, and cooperative control of multi-agent systems. 
\end{IEEEbiography}

\begin{IEEEbiography}[{\includegraphics[width=1in,height=1.25in,clip,keepaspectratio]{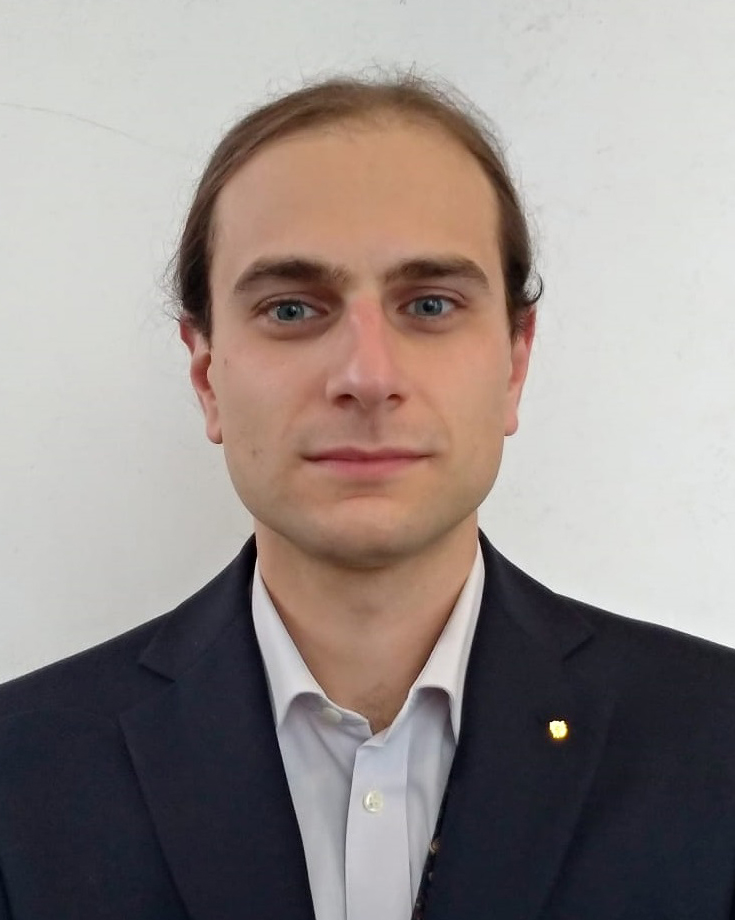}}]{Lorenzo Zino}  (M'21) has been an Assistant Professor with the Department of Electronics and Telecommunications at Politecnico di Torino (Turin, Italy) since 2022. He received the BS and MS in Mathematical Engineering from Politecnico di Torino, Torino, Italy, in 2012 and 2014, respectively, and the PhD in Pure and Applied Mathematics (with honors) from Politecnico di Torino and Università di Torino (joint doctorate program), in 2018. He was a Research Fellow at Politecnico di Torino (2018--19) and the University of Groningen (2019--22) and a Visiting Research Assistant at New York University Tandon School of Engineering (2017--18 and 2019). His current research interests include modeling, analysis, and control of dynamics over complex networks, applied probability, network analysis, and game theory. Since 2021, he has been an Associate Editor of the \emph{Journal of Computational Science}.
\end{IEEEbiography}


\begin{IEEEbiography}[{\includegraphics[width=1in,height=1.25in,clip,keepaspectratio]{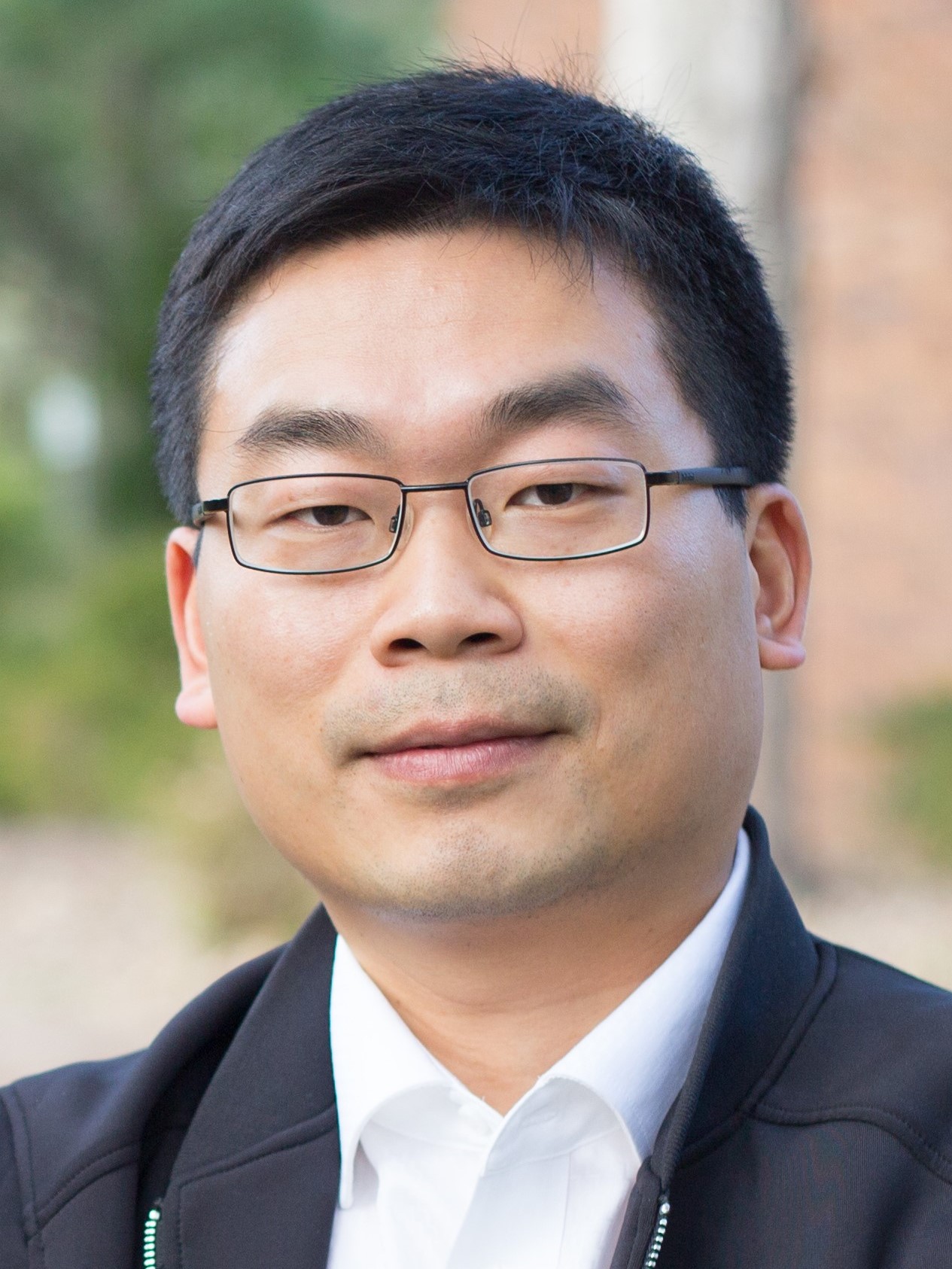}}]{Zhiyong Chen} received the B.E. degree from the
University of Science and Technology of China, and
the M.Phil. and Ph.D. degrees from the Chinese
University of Hong Kong, in 2000, 2002 and 2005,
respectively. He worked as a Research Associate
at the University of Virginia during 2005-2006. He
joined the University of Newcastle, Australia, in
2006, where he is currently a Professor. He was also
a Changjiang Chair Professor with Central South
University, Changsha, China. His research interests
include nonlinear systems and control, biological
systems, and multi-agent systems. He is/was an Associate Editor of Automatica, IEEE Transactions on Automatic Control and IEEE Transactions on
Cybernetics.
\end{IEEEbiography}

\begin{IEEEbiography}[{\includegraphics[width=1in,height=1.25in,clip,keepaspectratio]{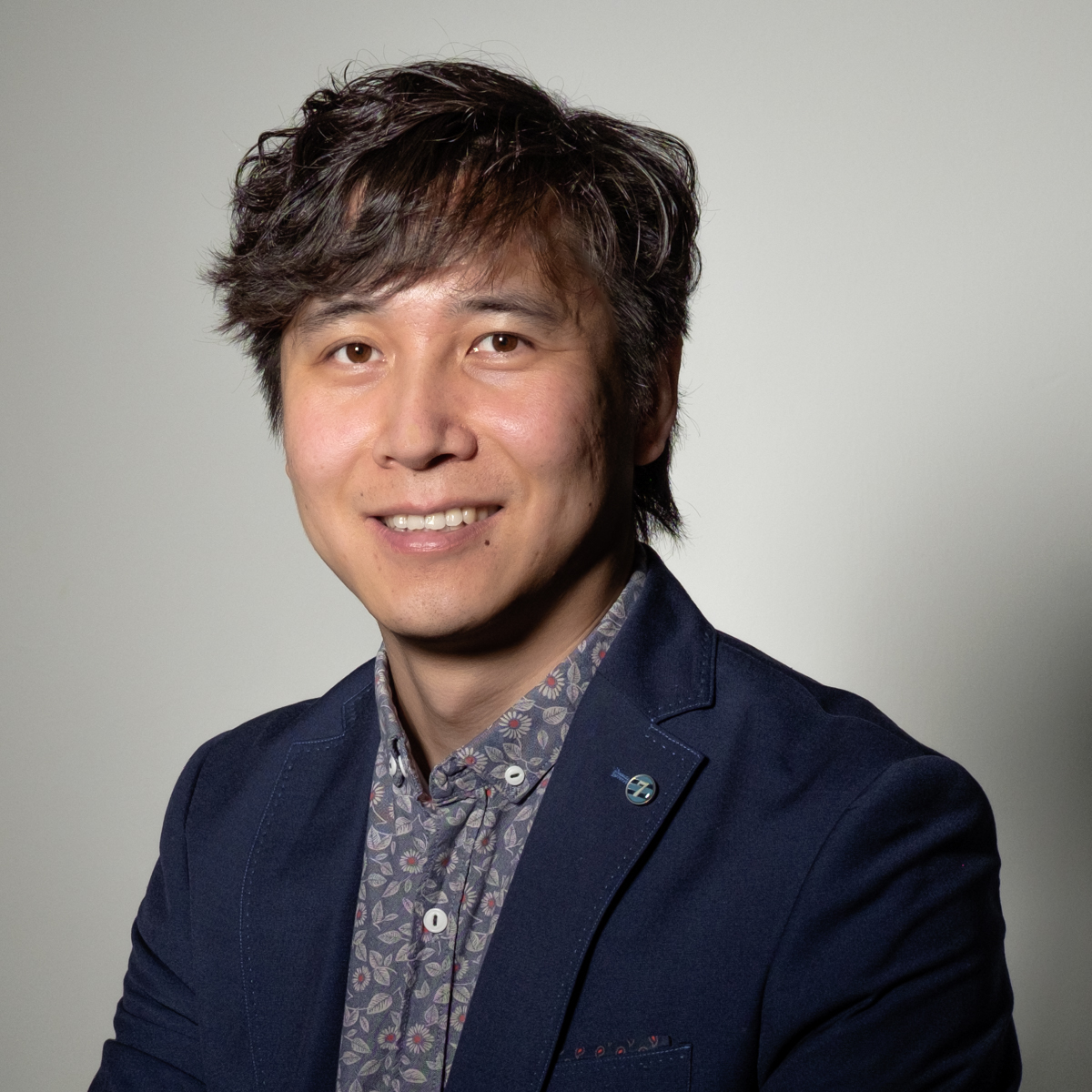}}]{Ming Cao} (S’05-M'08-SM'16-F'22) has since 2016
been a professor of networks and robotics with
the Engineering and Technology Institute (ENTEG) at the University of Groningen, the Netherlands, where he started as an assistant professor in 2008. He received the Bachelor degree in 1999 and the Master degree in 2002
from Tsinghua University, Beijing, China, and the
Ph.D. degree in 2007 from Yale University, New
Haven, CT, USA, all in Electrical Engineering.
From September 2007 to August 2008, he was
a Postdoctoral Research Associate with the Department of Mechanical
and Aerospace Engineering at Princeton University, Princeton, NJ, USA.
He worked as a research intern during the summer of 2006 with the
Mathematical Sciences Department at the IBM T. J. Watson Research
Center, NY, USA. He is an IEEE fellow. He is the 2017 and inaugural recipient of the Manfred Thoma medal
from the International Federation of Automatic Control (IFAC) and the
2016 recipient of the European Control Award sponsored by the European Control Association (EUCA). He is a Senior Editor for Systems and
Control Letters, an Associate Editor for IEEE Transactions on Automatic
Control, and was an associate editor for IEEE Transactions on Circuits
and Systems and IEEE Circuits and Systems Magazine. He is a member
of the IFAC Conference Board and a vice chair of the IFAC Technical
Committee on Large-Scale Complex Systems. His research interests
include autonomous agents and multiagent systems, complex networks, and decision-making processes.
\end{IEEEbiography}
\end{document}